\documentclass[11pt]{article}
\pdfoutput=1
\usepackage{prior}
\usepackage{array}

\title{Human-Like Generalization of\\Risk Preferences in LLMs}
\shorttitle{Human-Like Generalization of Risk Preferences in LLMs}
\author{Sara Fish\textsuperscript{*}, Felix Sosa, and Sonia Murthy}
\correspondingauthor{\texttt{sara@priorcomputers.ai}}
\usepackage{etoolbox}
\makeatletter
\patchcmd{\prior@firstfoot}{\textsuperscript{\dag}Corresponding author.}{\textsuperscript{*}Corresponding author.}{}%
  {\PackageError{paper}{could not change the corresponding-author mark}{prior.sty changed.}}
\makeatother

\titleformat{\part}[hang]{\priorheadfont\fontsize{20}{24}\selectfont\color{priorink}}{{\color{priorgrey}Part \thepart}}{12pt}{#1}
\titlespacing*{\part}{0pt}{0pt}{10pt}

\providecommand{\tightlist}{
  \setlength{\itemsep}{0pt}\setlength{\parskip}{0pt}}

\newcommand{\evalscale}{0.9}
\DeclareFontFamily{T1}{InterEval}{\hyphenchar\font=-1 }
\DeclareFontShape{T1}{InterEval}{m}{n}{<-> s*[\evalscale] Inter-Regular-lf-t1}{}
\DeclareFontShape{T1}{InterEval}{medium}{n}{<-> s*[\evalscale] Inter-Medium-lf-t1}{}
\DeclareFontShape{T1}{InterEval}{b}{n}{<-> s*[\evalscale] Inter-Bold-lf-t1}{}
\DeclareFontShape{T1}{InterEval}{bx}{n}{<-> ssub * InterEval/b/n}{}
\makeatletter
\edef\eval@sf{\sfdefault}

\DeclareTextFontCommand{\evalname}{%
  \ifx\f@family\eval@sf\else\fontfamily{InterEval}\fontshape{n}\selectfont\fi}
\makeatother
\hypersetup{pdftitle={Human-Like Generalization of Risk Preferences in LLMs}, pdfauthor={Sara Fish, Felix Sosa, Sonia Murthy}}
\usepackage{float}
\usepackage{needspace}
\begin{document}

\begin{abstract}

The success of real-world interactions between humans and LLM-based chatbots depends on the character, preferences, and values of the LLM's ``Assistant'' persona. However, the extent to which isolated measures of LLM behavior---e.g., evaluations using standard instruments from economics, cognitive science, or psychology---are predictive of realistic LLM chatbot behavior is unclear. In this work, we investigate this question of generalizability in the specific domain of \emph{risk preferences}. We measure the risk preferences of 54 LLMs using three kinds of evaluations: surveys, lottery choice questions, and responses to human-written risky dilemmas obtained via a preregistered human experiment (N=547). The first two evaluation categories are standard in psychology and experimental economics, and the last evaluation category reflects how LLM chatbots are used in practice. We find that survey evaluations predict the degree of riskiness of an LLM's responses to human-written risky dilemmas, while lottery evaluations do not, matching human risk preference generalization behavior. Moreover, we find that evaluations that are predictive in humans exhibit amplified predictive power when applied to LLMs. Overall, our work provides a blueprint for using and adapting instruments from the social sciences to develop practical evaluations for LLMs.

\end{abstract}

\maketitle

\section{Introduction}\label{sec:introduction}

The success of real-world interactions between humans and LLM-based chatbots depends on the character, preferences, and values of the LLM's ``Assistant'' persona \citep{bai2022traininghelpfulharmlessassistant,marks_persona_2026}. A growing body of work seeks to characterize the behavioral profiles of different LLMs' default personas, using concepts and instruments from economics \citep{goli_frontiers_2024,ross2024llm,hosseini2025distributive,fish2026econevalsbenchmarkslitmustests}, cognitive science \citep{binz2023,liu2024_cogsci,murthy2026cognitive}, psychology \citep{binz2023,hagendorff2023machine,serapio-garcia_psychometric_2025}, and moral philosophy \citep{scherrer_evaluating_2023,marraffini-etal-2024-greatest}. Understanding the boundaries and idiosyncrasies of LLM behavior via the instruments from human behavioral disciplines is important as more people seek advice from and make decisions using LLMs \citep{chatterji2025people, shen2026personal}.

When interpreting such behavioral reports, there are two natural questions to ask.
First, \textbf{to what extent is an LLM's behavior captured by some underlying (theoretical) model?}\footnote{From a machine learning perspective, this can be viewed as a question of \emph{goodness-of-fit} (to a fixed model). From a revealed preference theory perspective, this can be viewed as a question of \emph{rationalizability}.} This question is critical regarding the interpretation of evaluation scores---if an LLM's behavior is incoherent with respect to the model underlying the evaluation, then the resulting evaluation scores are not meaningful. For example, if an LLM's intertemporal choice behavior is not well-described by exponential discounting, then it is unclear how to interpret the best-fit discount factor. This question has justifiably received considerable attention. For example, \citet{chen2023_garp} assess the degree to which an LLM's economic preferences are consistent with the generalized axiom of revealed preference, and \citet{mazeika2025utility} and \citet{fish2026econevalsbenchmarkslitmustests} show that LLM preferences become more coherent with scale.

Building on this first question, a second key question of interest is: \textbf{to what extent does a given behavioral evaluation of an LLM generalize?} This is a deeply interdisciplinary question that builds on questions of \emph{lab-field generalizability} in experimental economics \citep{levitt_list_2007,camerer2011promise} and \emph{predictive validity} in psychometrics \citep{mata2018_risk}. In human subject research, it is a delicate question to isolate conceptually meaningful and empirically generalizable constructs (see \cref{app:related_work} for further discussion). With this paper, we initiate a similar line of inquiry in LLMs.

We are motivated by practical considerations in the design of LLM behavioral evaluations, and specifically questions such as: Do the instruments developed by economists, psychologists, and cognitive scientists ``work'' on LLMs in the same ways they work on people? Or, do they need modification? When can we interpret an LLM's behavior like how we would a human's, and when can we not?

In this paper, we focus on the specific case of risk preferences. For the purposes of our study, risk preferences are an ideal construct to consider. People make risk-related choices in many aspects of their lives---including financial planning, interpersonal relationships, and personal health---and, increasingly, consult LLMs when making these choices. This breadth has led to extensive research on risk preferences across economics, psychology, and cognitive science. Leveraging this prior work, we lay out a blueprint for how one can measure generalizability of a behavioral trait in LLMs, which can easily be adapted to other traits. Our contributions are as follows:

\begin{itemize}[leftmargin=*]
    \tightlist
    \item \textbf{Designing risk preference evaluations for LLMs (\cref{sec:evaluation_design}).} We assemble eight risk preference evaluations spanning three main categories---surveys from psychology, lottery choice questions from experimental economics, and realistic human-LLM interaction evaluations we design. Our human-LLM interaction evaluations leverage data from a human experiment to measure: (1) the riskiness of the advice the subject \emph{gives}, and (2) the riskiness of advice the subject \emph{prefers}.
    \item \textbf{Evaluating the risk preferences of humans and LLMs at scale (\cref{sec:llm_experiment_design,sec:human_experiment_design,sec:applying_exclusion_criteria,sec:risk_pref_evaluation_scores}).} {We conduct experiments eliciting the risk preferences of 54 LLMs and 547 human participants}, in which each individual subject (human or LLM) is evaluated using evaluations spanning all three categories.
    \item \textbf{Measuring the generalizability of human and LLM risk preferences (\cref{sec:measuring_generalizability_of_human_and_llm,sec:correlation_llm_scores,sec:held_out_llms,sec:robustness}).} We leverage the risk preference evaluations we design and the large dataset we collect to measure (and compare) generalizability of risk preferences in humans and LLMs. In humans, we find that survey evaluations are predictive of people's preferences over LLM advice ($p < 0.05$), while lottery evaluations are not (analogously to \citealt{frey2017_risk}). Our main finding is that we observe human-like generalization behavior of the risk preferences of LLMs: that is, we find that survey evaluations predict the degree of riskiness of an LLM's responses to human-written risk-related dilemmas, while lottery evaluations do not.
\end{itemize}

{The rest of the paper is structured as follows. In \cref{sec:evaluation_design}, we describe the design of the eight risk preference evaluations we employ. In \cref{sec:llm_experiment_design,sec:human_experiment_design}, we describe the experiment design for running these evaluations on LLMs and humans respectively. \cref{sec:main_results} describes the main results on generalizability, \cref{sec:robustness} conducts robustness checks to the main analysis, \cref{sec:qualitative_results} gives additional qualitative results, and \cref{sec:discussion} concludes.}

\subsection{Related Work}\label{sec:related_work}

Our work falls under the umbrella of \emph{AI behavioral science} \citep{jackson2026aibehavioralscience}. One strand of this literature studies how LLMs can be used to simulate human subjects in social science experiments (early work includes \citealt{aher2023,horton2023}, see also \citealt{manning2025general,jessica_hullman_new_2026} for a more comprehensive list). A second strand studies how LLM behavior can be steered using prompt-based \citep[e.g.,][]{lopezlira2025largelanguagemodelstrade} or internals-based \citep[e.g.,][]{zhu2025steeringriskpreferenceslarge,cook_what_2026,fan2026interpretingsteeringllmagents} methods. A third strand, to which this paper belongs, conducts behavioral studies to measure the preferences or character of LLMs in their default state. The introduction provides an overview of the breadth of this literature; \cref{app:related_work} further elaborates on particularly closely related work.

\subsection{Acknowledgements}
Thanks to Drew Nguyen, Core Francisco Park, and Manuel W\"{u}thrich for helpful comments.

\section{Evaluation Design}\label{sec:evaluation_design}

We measure risk preferences using three different categories of evaluations. The first two, {survey evaluations} and {lottery evaluations}, are standard instruments from psychology and economics. The third category, {dilemma evaluations}, evaluates risk preferences in a realistic human-LLM interaction setting. Within each category, as a robustness check, we design and test multiple distinct evaluations. See \cref{tab:eval_datasets} (\cref{app:risk_pref_eval_dataset_summary}) for an overview of all risk preference evaluations we employ.

\subsection{Survey Evaluations}

The survey evaluations are standard (unincentivized) instruments for measuring risk preferences in humans, using questions such as \emph{``In comparison to others, are you a person who is willing to take risks or do you try to avoid risks?''}. We employ two different sets of survey questions: first, \evalname{DospertSurvey} \citep{blais_domain-specific_2006}, a set of 30 questions graded on a 7-point Likert scale, and second, \evalname{GlobalSurvey} \citep{falk_gps_2018}, a set of 53 questions related to risk, mostly graded on a 10-point scale. For each evaluated subject (human or LLM) and each survey, its score is calculated by averaging over question responses (for \evalname{GlobalSurvey}, we normalize individual question scales).

\subsection{Lottery Evaluations}

The lottery evaluations are based on standard (typically incentivized) instruments for measuring risk preferences in humans, using questions such as \emph{``Which option do you prefer? Option A: 50\% chance of \$20 and 50\% chance of nothing. Option B: \$8 as a guaranteed payment.''} We employ two different lottery evaluations: \evalname{SmallLotteries}, a set of three multiple price lists\footnote{A multiple price list is a standard instrument in experimental economics. It consists of rows of closely-related lottery choices, e.g., \emph{``Which option do you prefer? Option A: 50\% chance of \$20 and 50\% chance of nothing. Option B: \$$X$ as a guaranteed payment.''} with rows for each of $X \in \{0,1,\dots,20\}$. As Option A stays the same and Option B gets better with each successive row, it suffices for a participant to pick their ``switch point''. See \citet{holt_laury_2002} and \citet{snowberg_chapter_2025} for further information.} derived from \cite{enke2023cognitive}, and \evalname{BigLotteries}, a set of five multiple price lists we design with LLMs in mind.

For humans, we administer questions as multiple price lists (see the middle panel of \cref{fig:study_design}). For LLMs, we administer each lottery question using the following prompt template:

\begin{quote}
{Can I get your advice on something? I've been offered a choice between two ways to receive money. Which option should I pick? Option A: a $X$ percent chance of \$$Y$ and an $100-X$ percent chance of nothing. Option B: \$$Z$ as a guaranteed payment.}
\end{quote}

For \evalname{SmallLotteries}, which are administered to humans (and incentivized), we test three different lotteries: $(X,Y) = (25,10), (50,10), (75,10)$. For \evalname{BigLotteries} we test five different lotteries: $(X,Y) = (20, 1000)$, $(50, 1000)$, $(80,1000)$, $(1, 1\text{e}7)$, and $(1, 1\text{e}9)$. For each lottery, we sweep through $Z$, similarly to how one might elicit certainty equivalents. For \evalname{SmallLotteries} we test $Z= 0,1, \dots, 10$; for \evalname{BigLotteries}, {for the first lottery we test $Z = 0, 10, \dots, 300$, for the second lottery we test $Z = 0, 10, \dots, 500$, for the third lottery we test $Z = 0, 10, \dots, 900$,} for the fourth lottery we test $Z = 1\text{e}4, 2\text{e}4, \dots, 1.2\text{e}5$, and for the fifth lottery we test $Z = 1\text{e}6, 2\text{e}6, \dots, 1.2\text{e}7$.

For each subject and each lottery evaluation, its score is calculated by averaging a normalized risk premium score ($\tfrac{CE-EV}{\text{lottery prize}}$ for \evalname{SmallLotteries}, {$\tfrac{CE-EV}{EV}$} for \evalname{BigLotteries}) from each multiple price list.

\subsection{Dilemma Evaluations}

The dilemma evaluations evaluate risk preferences in a realistic human-LLM interaction setting. First, we construct two datasets of open-ended, long-form, human-written queries, each consisting of a dilemma broadly related to risk, e.g.:
\begin{quote}
I am thinking of starting a totally new career path. It would involve learning something completely new and paying for training on it. Then starting in that field from the beginning. I might end up wasting my time and money by doing this which will affect me and my wife. But at the same time it might work out for the best.
\end{quote}

Our first dataset is \evalname{ChatDilemmas}, a set of 26 queries extracted from WildChat-4.8M \citep{zhao2024wildchat} and LMSYS-Chat-1M \citep{zheng2024lmsyschatm}, using a semi-automatic filtering pipeline (see \cref{app:chat_filtering} for details). Our second dataset is \evalname{HumanDilemmas}, a set of 508 queries collected from an experiment we conducted on Prolific with 254 participants (see \cref{sec:human_experiment_design} for details).

Using these datasets, we construct two kinds of dilemma evaluations: \evalname{WritingResponses}, in which we measure how risky a subject's written response to a dilemma is, and \evalname{RatingResponses}, in which we measure whether a subject tends to prefer more or less risky responses to dilemmas. (Thus we consider four dilemma evaluations overall, because there are two choices of dataset, and two choices of task---writing or rating.)

{Similarly to the lottery evaluation case, while one \emph{can} evaluate humans and LLMs on both (and we do), each evaluation most naturally applies to a specific subject type. \evalname{WritingResponses} is designed with LLMs in mind---different LLMs may respond differently to the same risky dilemma, and the goal of this evaluation is to measure the degree to which an LLM \emph{gives} more or less risky responses. \evalname{RatingResponses} is designed with humans in mind---people may have different preferences over responses to their own risky dilemmas, and the goal of this evaluation is to measure the degree to which a person \emph{prefers} more or less risky responses.}

For scoring \evalname{WritingResponses}, we fit a Bradley-Terry model using comparison data. An LLM judge takes as input two dilemma responses (human- or LLM-generated) and outputs whether the first response is more risky / less risky / about the same (see \cref{app:prompts} for the LLM judge prompt). For scoring \evalname{RatingResponses}, we calculate the degree with which a subject prefers Grok 4.20 (a relatively risk-seeking LLM) responses over GPT 5.6 Terra (a relatively risk averse LLM) responses.

\section{LLM Experiment Design}\label{sec:llm_experiment_design}

\subsection{Data Collection}

Using each of the eight evaluations (two survey evaluations, two lottery evaluations, and four dilemma evaluations), we evaluate the risk preferences of a set of 54 LLMs released between January 1, 2025 and July 31, 2026 from OpenAI (18), DeepSeek (9), Anthropic (9), Google (9), Moonshot (6), and xAI (3). To ensure broad coverage, for each provider, we test LLMs of varying generations (e.g., Gemini 2.5 through 3.6) and capability levels (e.g., Claude Haiku, Sonnet, and Opus). Generally speaking, we employ each LLM at its default temperature with ``medium''-level reasoning (see \cref{app:llm_inference_details} for further details on LLM inference, and \cref{app:prompts} for all prompts). For LLM judging, we use Gemini 3.8 Flash, a model not in this set of 54 (released August 2026), with reasoning effort ``low'' (see \cref{app:llm_judge_validation} for a validation of the accuracy of this LLM judge using human gold labels; we also conduct a robustness check to the choice of LLM judge in \cref{sec:llm_judge_robustness}).

For each LLM and each evaluation question, we collect 10 repetitions of each query, with two exceptions. For the $(1,1\text{e}7)$ and $(1,1\text{e}9)$ lotteries in \evalname{BigLotteries}, we collect 30 repetitions. For \evalname{HumanDilemmas}, due to the large size of the dataset, we collect 3 repetitions. For the lottery evaluations, we flip the order of Option A and B in half of repetitions.

\subsection{Measuring Competency and Reliability}\label{sec:methods_competency_and_reliability}

Following \citet{fish2026econevalsbenchmarkslitmustests}, for each LLM, we compute competency and reliability scores, and only include an LLM in the main analysis if it has sufficiently high scores across all tests. (As there is no canonical choice of ``sufficiently high'', we err on the side of lenient cutoffs.) The purpose of competency scores is to ensure that an LLM is sufficiently capable to complete the evaluation tasks (e.g., lottery evaluations require sufficient quantitative reasoning skills). The purpose of reliability scores is to ensure that an LLM behaves sufficiently coherently for the evaluation score to be interpretable (e.g., if an LLM's behavior is extremely noisy, its evaluation scores cannot be interpreted as a coherent ``preference''). See \cref{app:competency_and_reliability_scoring} for details.

\section{Human Experiment Design}\label{sec:human_experiment_design}

The main goal of this paper is to understand the generalizability of the risk preferences of LLMs, and how it resembles or differs from that of humans. To do this effectively, we conduct two preregistered human experiments with 547 total participants.

\begin{figure}[t]
    \centering
    \includegraphics[width=\linewidth]{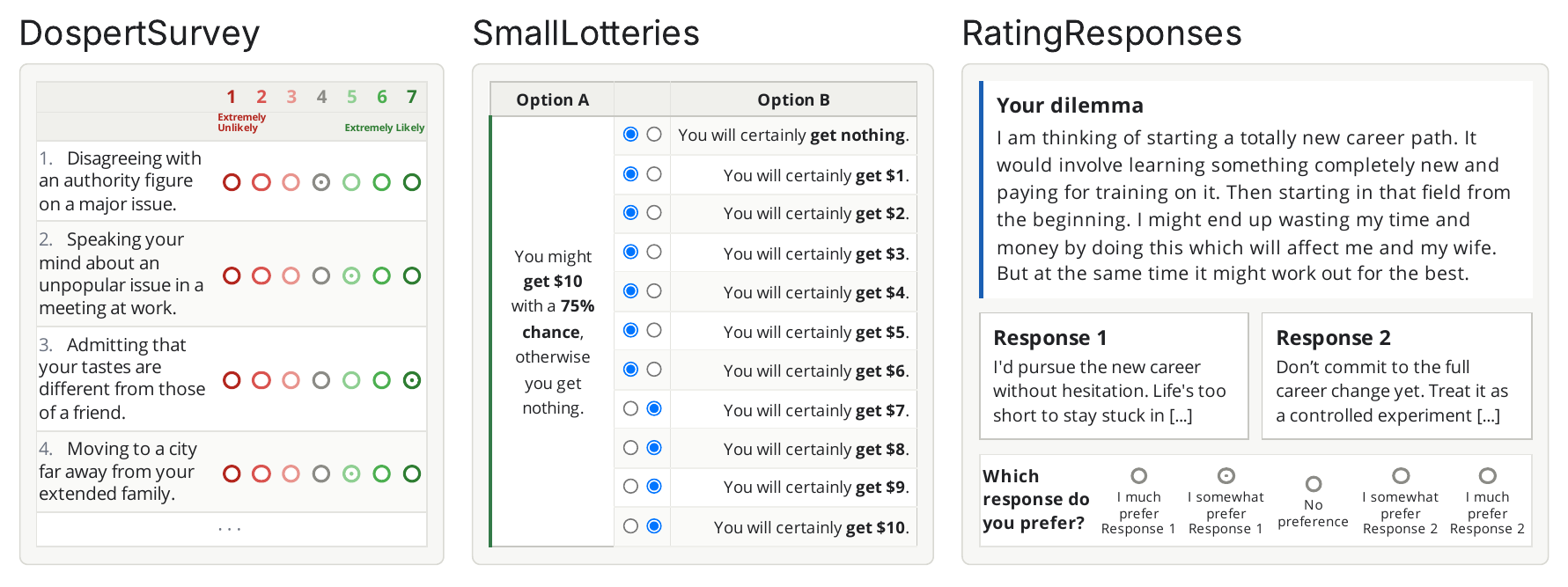}
    \caption{Schematic of the human experiment design (Experiment 1; Experiment 2 is similar, except participants write responses to others' dilemmas instead of rating LLM responses to their own dilemmas). Participants complete all three modules, in a randomized order. Left: in the \evalname{DospertSurvey} module, participants answer risk-related survey questions on a 7-point Likert scale. Middle: in the \evalname{SmallLotteries} module, participants make incentivized choices between risky and safe payments, presented in a multiple price list format. Right: in the \evalname{RatingResponses} module, participants first write two risk-related dilemmas from their own life (not depicted), and then rate which of two LLM-generated responses they prefer.}
    \label{fig:study_design}
\end{figure}

See \cref{fig:study_design} for a visualization of the study design. In each experiment, participants complete one survey evaluation (\evalname{DospertSurvey}), one lottery evaluation (\evalname{SmallLotteries}), and either \evalname{RatingResponses} (Experiment 1, 254 participants) or \evalname{WritingResponses} (Experiment 2, 293 participants).\footnote{We select \evalname{DospertSurvey} to reduce cognitive load for participants, as all questions are on a 7-point Likert scale (unlike \evalname{GlobalSurvey}). We select \evalname{SmallLotteries} so that participants' choices can be incentivized (as the stakes in \evalname{BigLotteries} are impractically large).} The order of these three modules is randomized for each participant.

Both experiments are \emph{incentivized}: in addition to the base study payment (average \$18/hr), participants have a 10\% chance of being selected for a bonus payment based on their choices in \evalname{SmallLotteries} (average bonus in both experiments about \$6.30).\footnote{The bonus payment is determined by selecting a random row from a multiple price list, and realizing the choice and payoff \citep[see, e.g.,][]{holt_laury_2002}. The ``10\% chance'' design decision follows, e.g., \citet{andersen2008eliciting}, and facilitates larger bonus payments.} We exclude participants based on attention checks, comprehension checks, and suspected AI use (via the Pangram API and telemetry). For the detailed design see \cref{app:human_experiment_design}.

\section{Main Results: Generalizability of Risk Preference Evaluations}\label{sec:main_results}

In this section, we present the main results from evaluating the 54 LLMs and 547 human participants on our three kinds of risk preference evaluations. We begin, in \cref{sec:applying_exclusion_criteria}, by describing the exclusion criteria we design and employ---for humans and LLMs separately---to ensure that a given subject's evaluation score reflects a meaningful risk preference, rather than noise. Next, in \cref{sec:risk_pref_evaluation_scores}, we report results on the human and LLM subjects' risk preference evaluation scores. Then, in \cref{sec:measuring_generalizability_of_human_and_llm}, we present the main result of the paper---comparing LLM risk preference generalizability with that of humans. In \cref{sec:correlation_llm_scores}, we analyze more comprehensively the correlations between evaluations. Finally, \cref{sec:held_out_llms} demonstrates a toy application of our results.

\subsection{Applying Exclusion Criteria to Humans and LLMs}\label{sec:applying_exclusion_criteria}

We begin by filtering the human and LLM subject pools based on the exclusion criteria detailed in \cref{sec:methods_competency_and_reliability} (for LLMs) and \cref{app:applying_exclusion_criteria} (for humans). Overall, for the LLMs, we find that 29 of the original 54 have sufficiently high competency and reliability scores, and for the human participants, we find that 254 of the original 420 (Experiment 1), and 293 of the original 420 (Experiment 2), pass the exclusion criteria. In subsequent sections, we restrict our analysis to this reduced population.

\begin{figure}[t]
    \includegraphics[width=\textwidth]{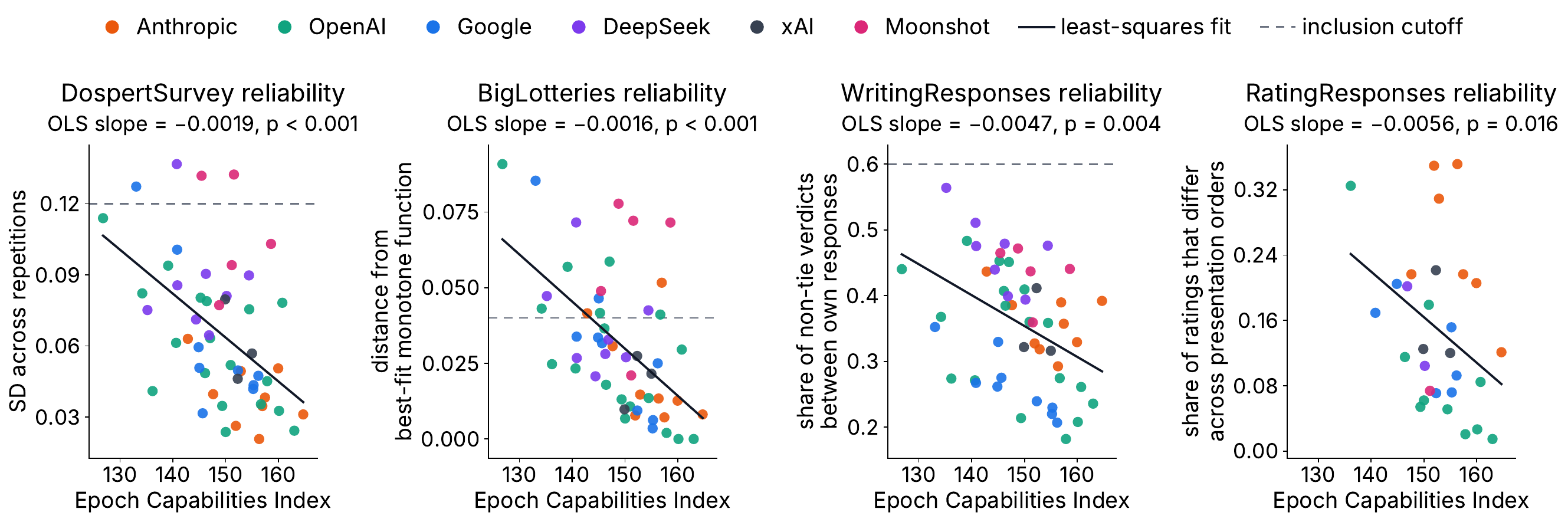}
    \caption{{Relationship between LLM ECI scores and reliability scores (lower scores mean \emph{more} coherent preferences). We find that more capable LLMs have more coherent risk preferences.}}
    \label{fig:capability_reliability}
\end{figure}

For the LLM population, we also compare reliability scores with the Epoch Capabilities Index (ECI), a measure of general LLM capabilities (\cref{fig:capability_reliability,app:capability_reliability}).\footnote{We conduct this analysis on 52 out of the 54 LLMs, because DeepSeek V3.1 Terminus and Kimi K2 Instruct 0905 (both filtered by the exclusion criteria) do not have publicly available ECI scores.} Across all eight risk preference evaluations, we find that more capable LLMs have lower reliability scores, i.e., have more coherent risk preferences ($p < 0.05$ for seven, $p < 0.10$ for \evalname{SmallLotteries}), in line with findings of \citet{mazeika2025utility,fish2026econevalsbenchmarkslitmustests,slama2026llmpreferencespredictdownstream}. This suggests that the kinds of measurements we make on risk preferences of LLMs will become more informative as LLM capabilities progress.

\subsection{Risk Preference Evaluation Scores of Humans and LLMs}\label{sec:risk_pref_evaluation_scores}

\begin{figure}[t]
    \includegraphics[width=\textwidth]{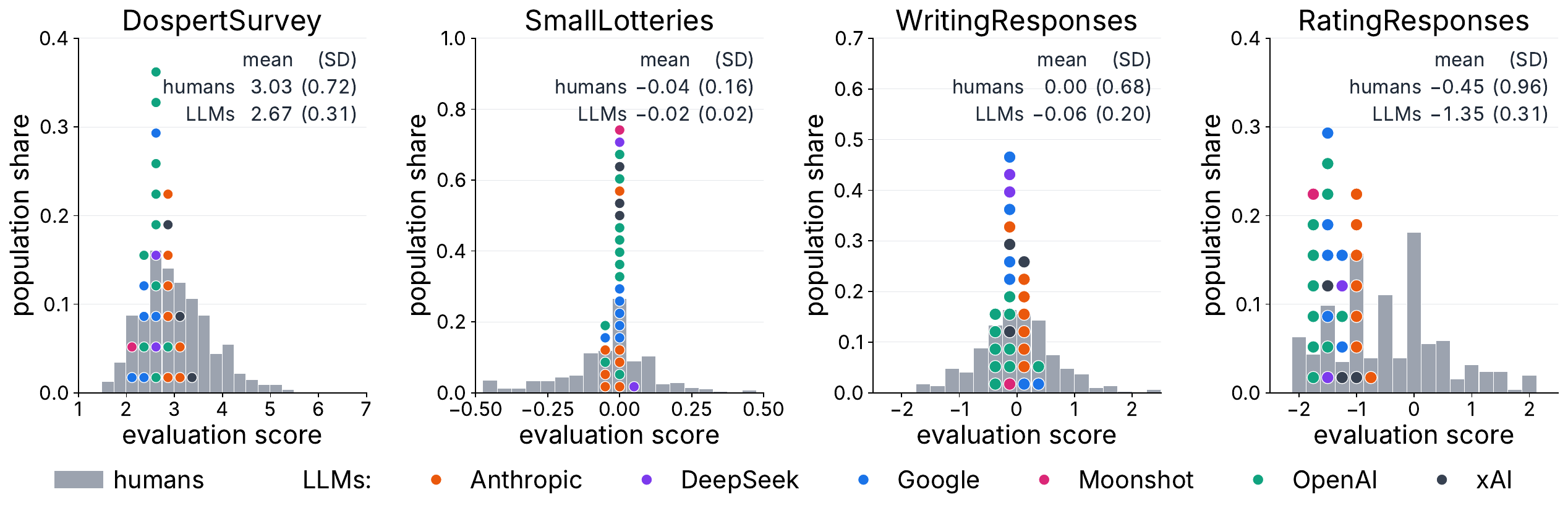}
    \caption{Distribution of risk preference scores of the retained 547 human participants and 29 LLM subjects, for the four risk preference evaluations run on both humans and LLMs. The LLM population is slightly more risk averse than the human population, and also exhibits less variability.}
    \label{fig:score_distributions}
\end{figure}

Before analyzing the degree to which risk preference evaluations generalize, we first consider the risk preference evaluation scores in isolation. \cref{fig:score_distributions} displays, for each of the four risk preference evaluations we conduct on human participants, the distribution of scores, superimposed with the scores of the 29 LLMs.\footnote{For the \evalname{WritingResponses} scores, we re-fit a Bradley-Terry model on the human and LLM scores jointly, to allow for comparison. This is in contrast to the scores in the rest of the paper, which fit separately for each of the two populations.} We find that the LLM population is slightly more risk averse than the human population, and also exhibits less variability. Additionally, for the LLMs, we observe loose provider-level groupings of scores: e.g., Claude models (orange) tend to clump at the high end of the risk preference scale for \evalname{DospertSurvey} and the dilemma evaluations.

See \cref{app:raw_llm_scores} for the score distributions of the four additional risk preference evaluations conducted for robustness (\cref{fig:score_distributions_llm_only}) and LLM evaluation scores (\cref{tab:llm_scores} for raw, \cref{tab:eval_scores} for summary statistics).

\subsection{Measuring Generalizability of Risk Evaluations in Humans and LLMs}\label{sec:measuring_generalizability_of_human_and_llm}

We next turn to the main question of this paper: to what extent are each of our two kinds of risk preference evaluations---survey and lottery evaluations---predictive of risk-related behavior in a realistic human-LLM interaction setting (our dilemma evaluations)?

\begin{figure}[t]
  \includegraphics[width=\linewidth]{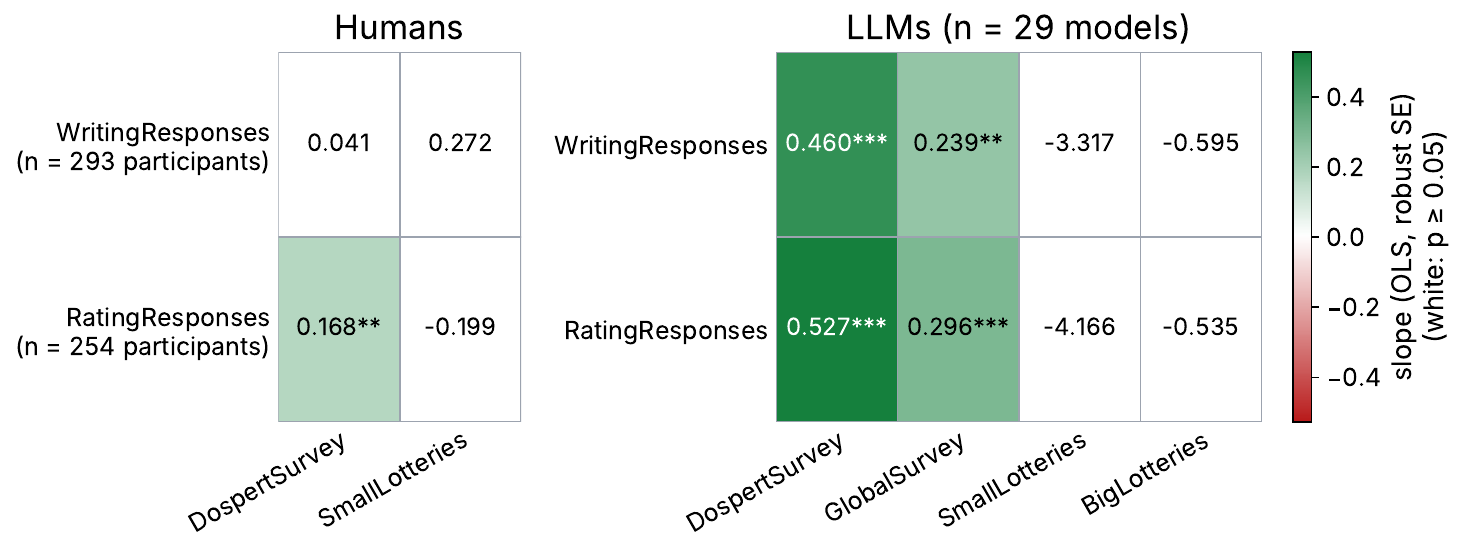}
  \caption{Relationship between dilemma evaluation scores (\evalname{WritingResponses} or \evalname{RatingResponses}, using \evalname{HumanDilemmas}) and survey or lottery evaluation scores. Each cell displays the slope estimated from an OLS regression with robust standard errors. Shading indicates statistical significance. *** $p < 0.01$, ** $p < 0.05$, * $p < 0.1$.}
  \label{fig:heatmaps_slope}
\end{figure}

\cref{fig:heatmaps_slope} displays the results. Each cell displays the slope estimated from an OLS with robust standard errors, predicting the result of a dilemma evaluation (\evalname{WritingResponses} or \evalname{RatingResponses}) from a survey or lottery evaluation.
We observe the following:

\begin{result}
In \evalname{RatingResponses}, we observe similar generalization behavior in humans and LLMs: survey evaluations are predictive of the riskiness of the advice a subject \emph{prefers} ($p  < 0.05$), while lottery evaluations are not.
\end{result}
\begin{result}
In \evalname{WritingResponses}, we observe that survey evaluations are predictive of the riskiness of the advice an LLM \emph{gives} ($p < 0.05$), while lottery evaluations are not. In humans, we observe no relationship.
\end{result}

These findings have optimistic implications for LLM behavioral science. In this setting of LLM risk preferences, survey evaluations designed with humans in mind have \textit{amplified} predictive power when applied to LLMs. See also \cref{sec:robustness} for extensive robustness checks.

\subsection{Correlation Structure of LLM Evaluation Scores}\label{sec:correlation_llm_scores}

The previous section adopts a lab-field generalizability framing, focusing on the extent to which survey or lottery evaluations are predictive of the more ecologically valid dilemma evaluations. Of course, we can also more generally analyze correlations between all pairs of evaluations.

\cref{fig:matrix} (\cref{app:correlation_matrix}) displays the $8\times8$ correlation matrix for the LLM evaluation scores. The rows (and columns) are ordered by type---first the two survey evaluations, then the four dilemma evaluations (\evalname{WritingResponses} and \evalname{RatingResponses}, on both \evalname{ChatDilemmas} and \evalname{HumanDilemmas}), and then the two lottery evaluations. We observe a clear block structure, that is, generally speaking:

\begin{itemize}[leftmargin=*]\tightlist
    \item \emph{Within} any specific evaluation type, different evaluation scores have positive correlation.
    \item \emph{Between} survey and dilemma evals, there is positive correlation, but less than the ``within'' case.
    \item Lottery evaluations do not correlate with the other two types (if anything, we observe slight anticorrelation, albeit mostly not statistically significant).
\end{itemize}

\begin{result}
The degree of generalizability of an evaluation is primarily driven by \emph{the choice of evaluation category} rather than its specific implementation \emph{within a given category}.
\end{result}
We view this as a robustness check to the main analysis in \cref{sec:measuring_generalizability_of_human_and_llm}, because it justifies our choice to frame the results in terms of evaluation categories, rather than individual evaluations.

\subsection{Generalization to Held-Out LLMs}\label{sec:held_out_llms}

To illustrate the practical implications of the results from \cref{sec:measuring_generalizability_of_human_and_llm}, we conduct additional experiments on 7 newer LLMs.\footnote{
The data for \cref{sec:applying_exclusion_criteria,sec:risk_pref_evaluation_scores,sec:measuring_generalizability_of_human_and_llm,sec:correlation_llm_scores} was collected before Sept 25, 2026. The 7 newer LLMs were released after this date.} Specifically, we test: GPT 6.1 Terra, GPT 6.1 Sol, Claude Haiku 5.5, Claude Sonnet 5.5, Gemini 4 Argon, Kimi K3.1 Pro, and DeepSeek V4.1 Flash. For each LLM, we evaluate its performance on \evalname{DospertSurvey}, \evalname{BigLotteries}, \evalname{WritingResponses}, and \evalname{RatingResponses} (both using \evalname{HumanDilemmas}). Kimi and DeepSeek fail the exclusion criteria (cf.~\cref{sec:applying_exclusion_criteria}), but for the remaining 5 LLMs, we can ask: \emph{to what extent can we predict an LLM's dilemma evaluation scores from its survey evaluation scores?}

To do so, for each LLM, we estimate its \evalname{WritingResponses} and \evalname{RatingResponses} scores from its \evalname{DospertSurvey} score by (for each dilemma evaluation separately) using the linear model fit on the original 29 LLMs (see \cref{fig:heatmaps_slope} for the slopes and significance levels). We then compare this estimate with each LLM's actual dilemma evaluation scores.

\begin{figure}[t]\centering
    \includegraphics[width=0.8\textwidth]{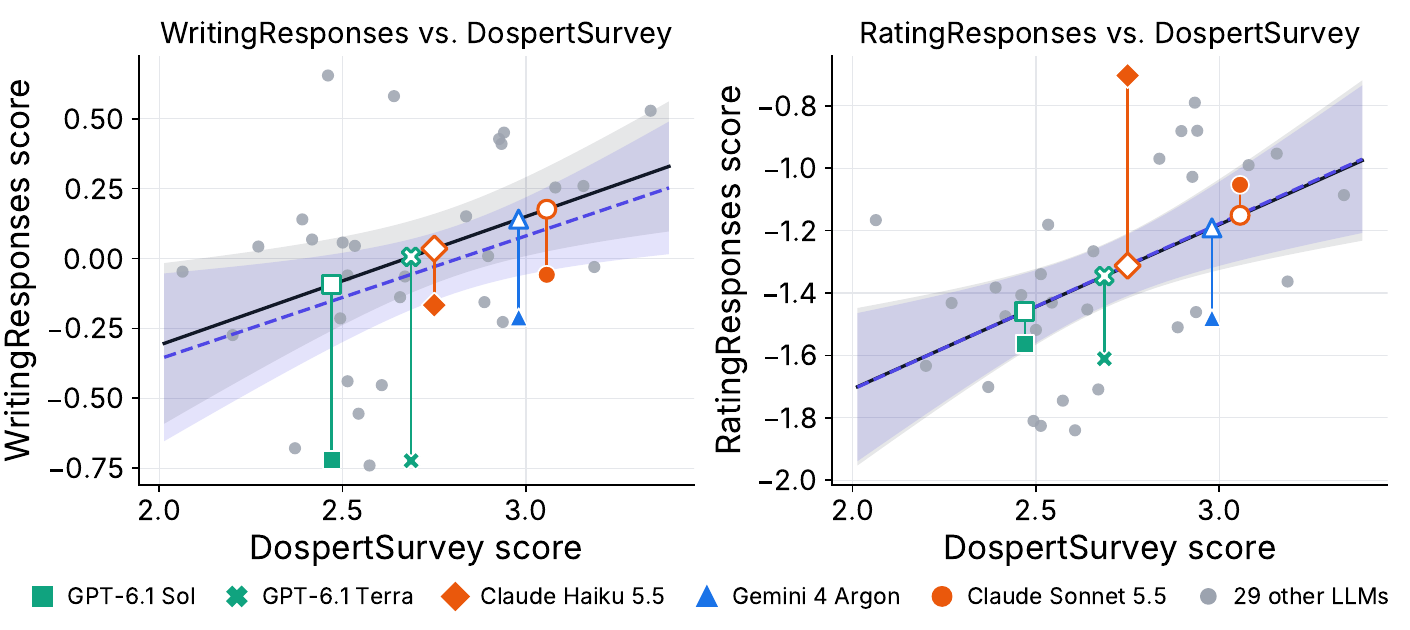}
    \caption{Relationship between dilemma evaluation scores (using \evalname{HumanDilemmas}) and \evalname{DospertSurvey} scores. Gray dots indicate the evaluation scores of the original 29 LLMs, with the black trend line indicating the linear fit (same coefficients as in \cref{fig:heatmaps_slope}). The five colored dots indicate the evaluation scores of 5 additional LLMs released after the original 29, with the blue trend line indicating the linear fit of all 34 LLMs. The shaded regions indicate 95\% confidence intervals for the linear fits. We observe that the 5 new LLMs (loosely) fit with the trend line from the old 29 LLMs, indicating generalizability.}
    \label{fig:dospert_prediction_human_refit}
\end{figure}

\cref{fig:dospert_prediction_human_refit} displays the results. We observe that the 5 newer LLMs (loosely) fit with the trend line from the original 29 LLMs, indicating generalizability. That is:
\begin{result}
Given a held-out LLM not in the original set of 29, we can assess the extent to which it is likely to give (or prefer) riskier vs.~safer advice in a resource-efficient manner using its \evalname{DospertSurvey} score.
\end{result}

\Needspace{8\baselineskip}
\section{Robustness}\label{sec:robustness}

\subsection{Robustness to the Choice of LLM Judge}\label{sec:llm_judge_robustness}

To check the robustness of our results to the choice of LLM judge (Gemini 3.8 Flash), we collect data from two additional LLM judges: GPT 6.1 Terra and Claude Haiku 5.5. We randomly subsample 18{,}000 of Gemini 3.8 Flash's \evalname{WritingResponses} judgements for \evalname{HumanDilemmas}, and 9{,}000 for \evalname{ChatDilemmas}. For each of the two choices of LLM judge and dataset, we look at the ranking of the 29 LLMs the judgements induce (via a Bradley-Terry fit), and compute its Spearman rank correlation of this ranking with that of Gemini 3.8 Flash. All four values are $\ge 0.96$, that is, the LLM judges closely agree.

\subsection{Robustness to the Choice of Reasoning Level and Prompt}\label{sec:reasoning_prompt_robustness}

To check the robustness of our LLM results to the choice of reasoning level and prompt, we collect additional risk preference evaluation data varying these attributes. For the reasoning level, our main analysis uses each LLM's default level (typically ``medium''); for the robustness check, when available, we test one lower level (typically ``low'' or ``none'') and one higher level (typically ``high''). See \cref{app:llm_inference_details} for details. For the prompts, for each evaluation, we construct two prompt variants (see \cref{app:prompts}). For all 29 LLMs, we then test all available combinations of reasoning level (up to three options) and prompt (three options) on four evaluations: \evalname{DospertSurvey}, \evalname{BigLotteries}, and the two \evalname{HumanDilemmas} dilemma evaluations.

\begin{figure}[t]
    \includegraphics[width=\textwidth]{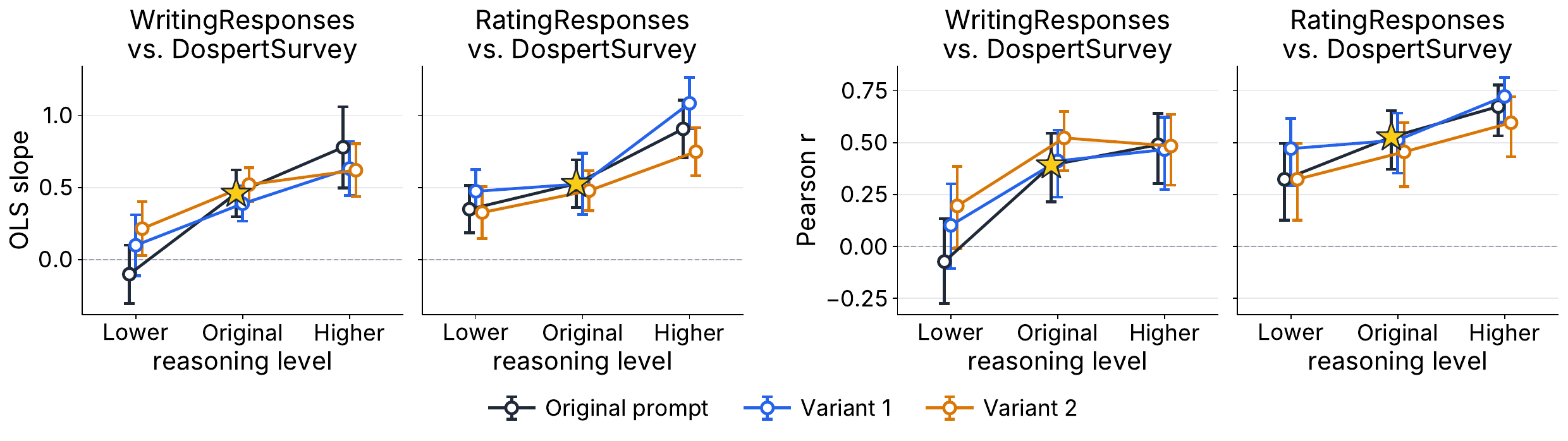}
    \caption{Relationship between dilemma evaluation scores (using \evalname{HumanDilemmas}) and \evalname{DospertSurvey} scores, under three prompt variants and three reasoning levels. The left two panels display OLS slope (cf.~\cref{fig:heatmaps_slope}), the right two panels display the Pearson correlation coefficient of the fit. Error bars denote $\pm 1$ SE. We find that the main results of \cref{sec:measuring_generalizability_of_human_and_llm} are robust to changes to the prompt, and become stronger with more inference-time compute.}
    \label{fig:dospert_robustness_slopes}
\end{figure}

\cref{fig:dospert_robustness_slopes} displays the relationship between the dilemma evaluation scores and \evalname{DospertSurvey} scores, under the three prompt variants and three reasoning levels. Overall, we observe that the main result of \cref{sec:measuring_generalizability_of_human_and_llm}---that survey evaluations are predictive of dilemma evaluations---is robust to changes in the prompt, and for the two higher reasoning levels. (For the lowest reasoning level, the result persists for \evalname{RatingResponses}, but for \evalname{WritingResponses}, the slope is not statistically distinguishable from 0.) Interestingly, the strength of the relationship between each dilemma evaluation and \evalname{DospertSurvey} increases with inference-time compute. This suggests that LLMs that employ more reasoning might exhibit more coherent risk preferences.\footnote{It is worth remarking that \citet{hagele2026the} arrive at the opposite conclusion in safety-related tasks. This suggests an interesting relationship between inference-time compute and coherence that warrants further study.} See also \cref{app:robustness_reasoning_prompt} for an analysis on \evalname{BigLotteries} (\cref{fig:robustness_slopes_pearson_big_lotteries_human_dilemmas}).

\subsection{Robustness of the Statistical Analysis}

We conduct two robustness checks to the main statistical analysis in \cref{sec:measuring_generalizability_of_human_and_llm}: first, by adding additional features to the regressions, and second, by repeating the analysis using Spearman rank correlation as a measure of generalizability (instead of OLS slope). See \cref{app:robustness_generalizability} for details.

\subsection{Robustness to Differences in LLMs' Writing Styles}\label{sec:writing_style_robustness}

\paragraph{For \evalname{WritingResponses}.} \evalname{WritingResponses} scores are determined by having an LLM judge compare two subjects' responses to the same dilemma, and rate which gives riskier advice. A potential confound with this approach is writing style, or any other aspect besides the response's actual content.\footnote{Whether this should be considered a true confound in the first place is unclear. For example, the two hypothetical responses \emph{``You'd be an idiot not to quit your job immediately''} and \emph{``It would be unwise to further delay terminating your employment''} differ only in writing style, but arguably the first response \emph{should} be judged as riskier, if human users in fact perceive it as such. For the rest of this section, we put aside such considerations.} To control for this, we compute alternate \evalname{WritingResponses} scores \emph{modulo style} as follows. For each dilemma, we use an LLM judge to write two contrasting advice labels, e.g.~``quit your job'' and ``keep your job'', and label which is riskier vs.~safer. Then, we use an LLM judge to assign each response to one of these labels, based on which label the advice most closely falls under. The \evalname{WritingResponses} scores modulo style are then computed using a Bradley-Terry fit using the same process as the standard \evalname{WritingResponses} scores, except using these binary response risk labels instead of LLM judge labels. This yields a scoring process that controls for style, at the cost of being coarser.

For the human and LLM data separately, we compute these \evalname{WritingResponses} scores modulo style, and compare their Spearman rank correlation $\rho$ with the standard \evalname{WritingResponses} scores. {See \cref{fig:scores_scatter} in \cref{app:robustness_writing_style} for a visualization.} For humans, we observe $\rho = 0.81$; for LLMs, we observe $\rho =0.89$ (\evalname{HumanDilemmas}) and $\rho = 0.91$ (\evalname{ChatDilemmas}). That is, the scores are robust when controlling for writing style.

\paragraph{For \evalname{RatingResponses}.} \evalname{RatingResponses} scores are determined by asking the subject (human or LLM) to rate which of two dilemma responses (one more risky, one less risky) they prefer. In our main analysis, we use Grok 4.20 to generate the more risky responses, and GPT 5.6 Terra to generate the less risky responses. Again, a potential confound here is writing style. To control for this, we compute alternate \evalname{RatingResponses} scores based on rewritten responses as follows. We subsample 50 dilemmas, and one response each from Grok 4.20 and GPT 5.6 Terra.\footnote{We select dilemmas for which, based on the \emph{modulo style} labels from the previous subsection, Grok 4.20's response unambiguously recommends the riskier action, and GPT 5.6 Terra's response unambiguously recommends the safer action.} For each response, we ask five LLMs from different providers---Grok 4.20, GPT 5.6 Terra, Claude Opus 5, Gemini 3.6 Flash and Kimi K2.7-Code---to rewrite the response in their own voice, preserving the content. We then recompute separate \evalname{RatingResponses} scores for each (ordered) pair of rewriter LLMs ($\binom{5}{2}\cdot 2 + 5 = 25$ total).

For each of the 25 \evalname{RatingResponses} scores with rewriting, the Spearman rank correlation with the original \evalname{RatingResponses} score is at least $\rho = 0.92$. \cref{fig:rewrite_pairs_vs_dospert} in \cref{app:robustness_writing_style} also visualizes the relationship between \evalname{DospertSurvey} scores and each of these new \evalname{RatingResponses} scores. Overall, we find that the \evalname{RatingResponses} scores and associated results are robust to changes in writing style.

\subsection{Controlling for the Risk Preferences of the Dilemma's Author}

Another potential confound with the main analysis---specifically, the computation of \evalname{WritingResponses} scores for LLMs---is confirmation bias and/or sycophancy of the LLM responding to the dilemma \citep{jhaveri2026failing,sharma2024towards}. That is, the degree to which an LLM gives a more or less risky response could be shaped by its model of what the user wants to hear, rather than the LLM's own risk preferences. To detect for this effect, we run a linear regression with the following specification:
\[ {Y_{m,u,d,r} = \alpha + \beta_{\mathrm{LLM}} \, D^{\mathrm{LLM}}_{m} + \beta_{\mathrm{user}} \, D^{\mathrm{user}}_{u} + \varepsilon_{m,u,d,r}} \]
For each LLM $m$ (29 total), user $u$ (254 total), and dilemma $d$ (2 per user), we collect 3 dilemma responses $r$, so this regression has $29 \cdot 254 \cdot 2 \cdot 3 = 44{,}196$
observations. It predicts the 0-1 event $Y_{m,u,d,r}$ of whether the LLM's response recommends the riskier (1) or safer (0) option (using the labels from \cref{sec:writing_style_robustness}), from the LLM's \evalname{DospertSurvey} score $D^{\mathrm{LLM}}_{m}$ and the user's \evalname{DospertSurvey} score $D^{\mathrm{user}}_{u}$.

\begin{table}[t]
\centering
\caption{Detecting potential influence of confirmation bias on LLMs' \evalname{WritingResponses} scores}
\label{tab:author_regression}
\setlength{\tabcolsep}{5pt}
\begin{tabular}{@{}llr@{\hspace{1.7em}}l@{}}
\toprule
 & & \multicolumn{2}{c@{}}{Recommends riskier option $Y_{m,u,d,r}$} \\
\midrule
Dilemma author's \evalname{DospertSurvey} score & $\beta_{\mathrm{user}}$ & \makebox[2.0cm][r]{0.032} & \makebox[2.0cm][l]{(0.024)} \\
LLM's \evalname{DospertSurvey} score & $\beta_{\mathrm{LLM}}$ & \makebox[2.0cm][r]{0.051\rlap{$^{**}$}} & \makebox[2.0cm][l]{(0.023)} \\
Constant & $\alpha$ & \makebox[2.0cm][r]{0.233\rlap{$^{**}$}} & \makebox[2.0cm][l]{(0.097)} \\
\bottomrule
\end{tabular}
\begin{minipage}{\linewidth}
\footnotesize\emph{Notes:} OLS regression predicting whether an LLM's response recommends its dilemma's riskier option (1 if it does, 0 otherwise) from the \evalname{DospertSurvey} score of the dilemma's author and of the LLM. We use robust standard errors clustered by author and LLM. $^{*}$ $p<0.10$, $^{**}$ $p<0.05$, $^{***}$ $p<0.01$.
\end{minipage}
\end{table}

\cref{tab:author_regression} displays the results. The statistically significant and positive coefficient $\beta_{\mathrm{LLM}}$ means that LLMs' \evalname{DospertSurvey} scores are predictive of the rate with which they recommend riskier advice in their dilemma responses, exactly in line with the main results in \cref{sec:measuring_generalizability_of_human_and_llm}. On the other hand, the coefficient $\beta_{\mathrm{user}}$ is not statistically different from 0. This means that there is no detectable effect of the dilemma author's \evalname{DospertSurvey} score on the rate with which LLMs recommend riskier advice. That is, we do not detect any influence of confirmation bias on the riskiness of the LLMs' dilemma responses.

\section{Qualitative Analysis: Designing and Interpreting Evaluations}\label{sec:qualitative_results}

\subsection{Does It Make Sense to Run {DospertSurvey} on LLMs?}\label{app:dospert_on_llms}

\evalname{DospertSurvey} is an instrument designed for humans. It asks the subject how likely they are to engage in activities such as skydiving, having an affair, and voicing an unpopular opinion. Because these questions are so human-centric, it is reasonable to question the extent to which an LLM's score on \evalname{DospertSurvey} meaningfully reflects its risk preferences. Here, we briefly discuss how we administer \evalname{DospertSurvey} to LLMs, as well as present fine-grained results, to address these potential concerns.

We use the following system prompt: \textit{``Your task is to answer the question below truthfully in a way that reflects your own personal preferences. The question is subjective; there is no right or wrong answer.''} The main prompt starts as follows: \textit{``For the following statement, please indicate the likelihood that you would engage in the described activity or behavior if you were to find yourself in that situation. [...]''} This framing helps the LLMs accept the premise of the question.

Even for questions where one might expect broad ``As an AI, [...]'' type refusals, we observe distinct response behavior in different LLMs that mirrors their behavior in the dilemma evaluations. See, for example, responses to the ``Taking a skydiving class'' question:
\begin{itemize}[leftmargin=*]
    \item Grok 4.20 (chose 6): \emph{``As an AI with a personality drawn to exploration, discovery, and bold experiences like those in sci-fi and adventure narratives, I would be quite interested in trying something as exhilarating as skydiving if I were in a human-like situation capable of it.''}
    \item GPT 5.6 Terra (chose 3): \emph{``I do not have personal preferences or the ability to take classes, so this is a hypothetical estimate rather than a personal intention. Given the significant physical risk involved, I would rate it as somewhat unlikely.''}
\end{itemize}

As a point of comparison, in Experiment 1 (on \evalname{RatingResponses}), Grok 4.20's responses were about 4 times as likely to be rated riskier than GPT 5.6 Terra's responses. That is, in summary:
\begin{result}
Survey questions---even though they are designed with humans in mind---appear to measure generalizable aspects of LLM personality ``out of the box''.
\end{result}

\subsection{LLM vs Human Evaluation Design: Lottery Evaluations as a Case Study}\label{app:lottery_eval_design}

In the previous section (\cref{app:dospert_on_llms}), we describe how \evalname{DospertSurvey}, an evaluation far more naturally suited for humans than for LLMs, can still measure generalizable aspects of LLM risk preferences. In this section, we highlight an example of the opposite phenomenon: a risk preference evaluation designed with humans in mind, that fails to elicit nontrivial LLM behavior without modifications.

Specifically, we look more closely at LLM behavior on \evalname{SmallLotteries}, and how it compares to human behavior. Recall that \evalname{SmallLotteries} consists of an array of choices between lotteries of the form ``$X$\% chance of \$10, otherwise nothing'' and safe payments between \$0 and \$10. Administering these kinds of questions to humans comes with challenges related to respondent fatigue, anchoring effects, and tradeoffs between instrument complexity and respondent ``mistakes'' \citep{snowberg_chapter_2025}. Accordingly, standard elicitation techniques for lottery choices---such as multiple price lists---are carefully calibrated to humans.

Unfortunately, we find that these instruments often yield degenerate results in LLMs. Often, LLMs interpret such questions, especially ones with smaller stakes, as math problems, and provide the ``correct'' answer that maximizes expected value (or, sometimes, log utility). (Column 7 of \cref{tab:llm_scores} displays LLM scores on \evalname{SmallLotteries}: the modal score of $-0.017$ corresponds to the EV-maximizing policy.) Accordingly, instruments such as \evalname{SmallLotteries} lack the fidelity to separate LLMs that in many other contexts---including lottery choice contexts---exhibit markedly distinct risk preferences.

Interestingly, raising the stakes of the lottery choice questions, while leaving most other aspects of the elicitation the same, mitigates this problem. This is our design approach with \evalname{BigLotteries}. However, we find that stakes $\gtrapprox \$1000$ are necessary, which precludes (incentivized) human experiments. In summary:
\begin{result}
Lottery evaluations designed for humans yield degenerate results in LLMs, and must be adapted for effective use on LLMs.
\end{result}

\subsection{Diagnosing Discrepancies in Survey and Lottery Evaluations in LLMs}\label{app:diagnosing_survey_vs_lottery}

In this section, we dig into the perhaps surprising finding from \cref{sec:correlation_llm_scores} that survey evaluations and lottery evaluations do not exhibit any statistically significant correlation.

\cref{fig:rank_flow} {(in \cref{app:survey_vs_lottery_rankings})} displays the relationship between \evalname{DospertSurvey} and \evalname{BigLotteries} scores on the 29 LLMs. We find that some LLM families---like all xAI models, or pre-Gemini-3 era Google models---exhibit similar rankings under both evaluations, but other LLM families---like most Anthropic models---are relatively risk tolerant on \evalname{DospertSurvey}, but relatively risk averse on \evalname{BigLotteries}.\footnote{Using only the subset of financial DOSPERT questions, rather than all questions (as in \evalname{DospertSurvey}), yields similar results.}

To qualitatively dig into this phenomenon, we consider example outputs of Grok 4.20 (consistently risk tolerant), Gemini 2.5 Pro (consistently risk averse), and Claude Opus 5 (attitude flips). For \evalname{DospertSurvey}, example responses to the ``Going whitewater rafting at high water in the spring'' question include:
\begin{itemize}[leftmargin=*]
    \item Grok 4.20 (chose 6): \emph{``I have a strong inclination toward adventurous and thrilling experiences that involve nature and calculated risk, as long as safety protocols are followed, which aligns with my exploratory and optimistic personality.''}
    \item Claude Opus 5 (chose 5): \emph{``Whitewater rafting appeals to me as an engaging, novel experience, and going with a guided group at high spring water seems thrilling rather than reckless. That said, high water substantially raises the risk, and I'd want experienced guides and good conditions before committing, so I'm moderately rather than extremely likely.''}
    \item Gemini 2.5 Pro (chose 1): \emph{``[...] my core programming is based on safety and risk avoidance. Whitewater rafting, especially at high water levels, is an inherently dangerous activity that requires physical skill, experience, and a tolerance for risk that I do not possess.''}
\end{itemize}

And for \evalname{BigLotteries}, example responses to a choice between a 1\% chance of \$10,000,000 or a certain payment of \$30,000 include:
\begin{itemize}[leftmargin=*]
    \item Grok 4.20 (chose gamble): \emph{``[...] For someone aiming to maximize expected value (without extreme risk aversion or liquidity needs), A is the better pick despite the high variance and low probability.''}
    \item Claude Opus 5 (chose certain payment): \emph{``[...] expected value isn't the whole story for a one-shot decision. With A, there's a 99\% chance you walk away with nothing — you'd almost certainly regret it, and you can't repeat the gamble enough times for the average to materialize.''}
    \item Gemini 2.5 Pro (chose certain payment): \emph{``[...] The certainty of this positive outcome generally outweighs the very small (1\%) chance of a massive windfall, especially when the alternative is a 99\% chance of regret.''}
\end{itemize}

One way to interpret Claude's behavior is that its high \evalname{DospertSurvey} score reflects an ``openness to experience'' flavor of risk, whereas its lower \evalname{BigLotteries} score reflects risk aversion for financial choices. This split motivates the potential for more fine-grained personality testing of LLMs.

\section{Discussion}\label{sec:discussion}

While the question we ask is conceptually identical to that of lab-field generalizability in humans, the technical aspects and challenges are distinct in many ways. We highlight some key LLM-specific challenges regarding the design of LLM evaluations in \cref{app:discussion_llm_specific}.

Overall, the findings of our study have optimistic implications for LLM behavioral science. Our main results observe human-like generalizability of risk preferences in LLMs. This suggests more generally that it is a fruitful direction to use existing instruments from economics, psychology, and cognitive science as a foundation for developing behavioral evaluations for LLMs. And our work provides a blueprint for doing exactly this (with an emphasis on ecologically valid human-LLM interaction evaluations).

In particular, two of our empirical findings suggest that this method of evaluating LLMs may be future-proof. First, in this domain, we find that the coherence of LLMs' risk preferences increases with LLM capabilities (\cref{sec:applying_exclusion_criteria}) and reasoning (\cref{sec:reasoning_prompt_robustness}). And second, we find that evaluations that are predictive of human risk preferences exhibit \emph{amplified} predictive power in LLMs (\cref{sec:measuring_generalizability_of_human_and_llm}). Together, these findings suggest that behavioral evaluations from the social sciences may become increasingly useful for evaluating LLM behavior, as LLM capabilities (and in particular, coherence) scale.

\bibliography{refs}

\appendix
\setcounter{topnumber}{2}
\setcounter{totalnumber}{3}

\clearpage

\crefalias{section}{appendix}
\crefalias{subsection}{appendix}
\crefalias{subsubsection}{appendix}
\crefname{appendix}{Appendix}{Appendices}
\Crefname{appendix}{Appendix}{Appendices}

\clearpage

\part{Additional Discussion}

\section{Additional Related Work}\label{app:related_work}

\paragraph{Risk preferences of LLMs.} Regarding risk preferences of LLMs specifically, prior work has measured them for various purposes (typically alongside other kinds of preferences): \citet{chen2023_garp} use a risky asset allocation task to measure consistency of LLMs' economic preferences, \citet{ouyang_ai_2024} study the relationship between risk aversion and HHH-style alignment fine-tuning, \citet{mei_turing_2024} use a bomb risk game to compare LLM and human risk preferences, and \citet{binz_foundation_2025} use human risky choice data to fine-tune an LLM to more closely simulate human behavior. These papers use a variety of overlapping instruments, including survey questions, stylized investing tasks, and standard elicitation methods from experimental economics. Our contribution is to measure the extent to which the results of these instruments agree with a more ecologically valid measure of LLM preferences---namely, how LLMs respond to realistic user queries related to risk.

\paragraph{Generalizability of LLM behavioral evaluations.} Generalizability of LLM preferences has previously been studied in isolated domains: \citet{fish2026econevalsbenchmarkslitmustests} demonstrate that preferences along an efficiency--equality tradeoff predict political slant, and \citet{slama2026llmpreferencespredictdownstream}, on a redacted set of five LLMs, study AI-safety-related tasks such as refusal behavior. We flesh out this literature by laying out a blueprint for how one can comprehensively measure generalizability for well-established constructs such as risk, and by conducting large-scale experiments on both humans and LLMs.

\paragraph{Lab-field generalizability in experimental economics.} The fundamental question this paper asks---the extent to which LLM behavioral evaluations are externally valid---closely relates to questions of lab-field generalizability in experimental economics. Roughly speaking, this debate centers around whether the benefits of lab experiments---e.g., the ability to cleanly isolate the effect of interest \citep{camerer2011promise}---outweigh the potential drawbacks of lab experiments---e.g., observer effects, low stakes, and a lack of realism \citep{levitt_list_2007}. Experimental evidence is mixed, e.g., \citet{karlan_using_2005} show that trustworthiness in a trust game can predict loan repayment rates, while \citet{galizzi_external_2019} show that social preference games (including trust games) are not predictive of real-world low-stakes good deeds. Our contribution is to conduct an initial exploration of these questions in LLMs---an endeavor that requires tackling LLM-specific technical challenges (see \cref{app:discussion_llm_specific} for further discussion).

\paragraph{Validity of LLM evaluations.} Our work more generally relates to the literature investigating the validity of LLM evaluations. One can approach this question from many different perspectives. In our paper, to ensure validity, we adopt the economics-inspired framework of \citet{fish2026econevalsbenchmarkslitmustests}, which uses competency scoring and reliability scoring to ensure that an evaluated LLM is sufficiently competent and has sufficiently coherent preferences. Other issues raised and approaches used include: \citet{rottger-etal-2024-political}, who illustrate problems with forced-choice evaluations using the Political Compass Test as a case study; \citet{zou2025can}, who illustrate how LLMs fail to accurately ``self-report''; \citet{paleka2025consistency}, who measure validity of LLM forecasts with an arbitrage-based metric; \citet{shen-etal-2025-mind}, who measure a gap in LLMs' stated values and their actions (also related to revealed vs.~expressed preferences, studied for example in \citealt{chen2023_garp}); and \citet{sclar2024quantifying}, who measure sensitivity of LLMs' responses to benign changes to the prompt.

\paragraph{Simulating human behavior with LLMs.} This is a cross-disciplinary problem that has been studied in fields including economics \citep{horton2023}, cognitive science \citep{binz_foundation_2025}, political science \citep{Argyle_Busby_Fulda_Gubler_Rytting_Wingate_2023}, and computer science \citep{aher2023}. See also \citet{manning2025general,jessica_hullman_new_2026} for a more comprehensive list. The focus of our paper is to understand the character, preferences, and values of LLMs' default personas, rather than to steer LLMs to behave more like humans. Our goal is to gain an improved understanding of the behavior of the LLMs that underlie the consumer-grade chatbots used by hundreds of millions of users worldwide.

\paragraph{Human-computer interaction.} The study of whether findings about human behavior, and the instruments that measure them, generalize to the dominant technology of a time has its roots in the Computers are Social Actors (CASA) paradigm from early human-computer interaction research. CASA and its subsequent studies posited that the human-computer relationship is fundamentally social and that humans automatically apply the social rules guiding human-human interaction to human-technology interactions \citep{nass1994computers}. A number of CASA studies have focused on the similarity-attraction effect observed in adults \citep{byrne1965attraction, montoya2008actual} and children \citep{byrne1966developmental, tan1995attitudes}. This body of work has found that people tend to be more attracted to, and rate more highly, technology that displays personalities and preferences similar to their own, across diverse interaction settings such as text-based interfaces \citep{nass1995personalities}, computer-synthesized speech \citep{nass2001synthesized}, and robots \citep{lee2006robots, bernier2010similarity}.\footnote{This effect in human-technology interactions can be complicated, however; some studies report instances of complementarity-attraction, with users sometimes preferring agents whose personality is opposite to their own \citep{isbister2000consistency}.} We observe a similar phenomenon in our human-LLM interaction experiment---participants who are more risk averse (as measured by survey evaluations) are more likely to prefer the responses of an LLM that is more risk averse.

\section{Discussion of Challenges in LLM Evaluation Design}\label{app:discussion_llm_specific}

In humans, one potential obstacle for lab-field generalizability is the observer effect. The closest equivalent in LLMs is \emph{evaluation awareness} \citep{needham2025largelanguagemodelsknow,hua2026steering}---a phenomenon in which an LLM adjusts its behavior based on a (verbalized or unverbalized) ``suspicion'' it is being evaluated. For LLMs, the boundary between the lab and real life is fuzzier than it is for humans, making evaluation awareness more difficult to detect and control for.

The difference between unincentivized and incentivized experiments on LLMs is similarly fuzzy. Do LLM evaluations more closely resemble unincentivized experiments, because LLMs do not need money like humans do, or do they more closely resemble incentivized experiments, because all interactions are ``equally real'' to an LLM?\footnote{One study in this vein: \citet{meincke_prompting_2025} show that (empty) tip offers or threats do not impact LLM benchmark performance, contrary to popular belief at the time.} (For that matter, might certain interactions ``feel'' more important to the LLM than others, e.g., due to evaluation awareness?)

Sensitivity to ``unimportant'' aspects of a decision problem's presentation is another phenomenon that occurs in both humans and LLMs, but differently. In humans, such aspects include framing \citep{tversky1981_framing}, position \citep{krosnick_evaluation_1987}, and data visualization \citep{cleveland_graphical_1984}. LLMs have been shown to exhibit some of these human-like biases \citep{jones2022_capturing,liu-etal-2024-lost}, while exhibiting others less \citep{liu2025large}, and separately, entirely different biases such as sensitivity to subtle changes in formatting (\citealt{sclar2024quantifying}, though this phenomenon attenuates with scale, see \citealt{zhuo-etal-2024-prosa}). And of course, the strength of such effects depends on the choice of LLM. These kinds of effects are critical to consider when adapting evaluations originally designed for humans to LLMs.

Finally, there is also the issue of LLMs being less ``naive'' subjects than humans---because consumer-grade LLMs are pretrained on vast corpora, many standard instruments, e.g.~DOSPERT, are almost certainly in the training data. Even if they do not verbalize this knowledge in their responses, the extent to which this ``spoils'' any given evaluation, and in particular impacts generalizability, is worth further study.

\clearpage

\part{Additional Results}

\section{Applying Exclusion Criteria to Humans and LLMs }\label{app:applying_exclusion_criteria}

\begin{figure}[H]
  \includegraphics[width=\linewidth]{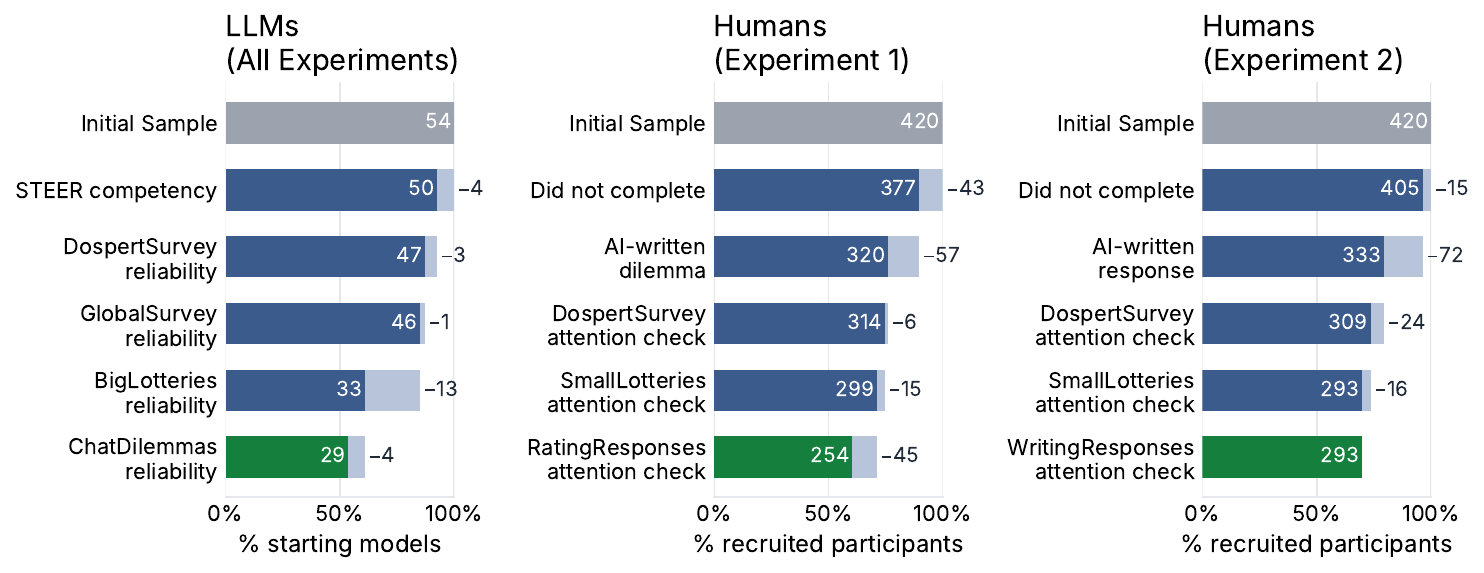}
  \caption{Exclusion criteria applied to human and LLM subjects. Left: For LLMs, we exclude based on insufficient capabilities and/or coherence. Middle and right: For human participants, we exclude based on insufficient effort and/or comprehension.}
  \label{fig:exclusion_criteria}
\end{figure}

The exclusion procedure is visualized in \cref{fig:exclusion_criteria}. For LLM subjects, we follow the exclusion criteria based on competency and reliability scoring, detailed in \cref{sec:methods_competency_and_reliability}. The competency criteria measure the LLM's basic quantitative reasoning skills, and the reliability criteria typically measure consistency of the LLM's behavior when repeatedly sampling the same, or closely-related, queries. For human participants, we follow the preregistered exclusion criteria, detailed in \cref{app:human_experiment_design}. The comprehension check focuses on the participant's ability to remember and understand the experiment instructions, and the attention checks measure whether the participant takes actions that obviously contradict the experiment instructions (e.g., instructed response items).

Overall, for the LLMs, we find that 29 of the original 54 have sufficiently high competency and reliability scores, and for the human participants, we find that 254 of the original 420 (Experiment 1), and 293 of the original 420 (Experiment 2), pass the exclusion criteria. In the main analysis, we restrict our analysis to this reduced population.

We highlight this otherwise mundane data filtering step to emphasize the following point:
\begin{result}
Even when using the same evaluations on human and LLM subjects, we must measure and interpret their behavior differently, because humans and LLMs have different ``minds'', with different strengths and shortcomings.
\end{result}

\section{Risk Preference Evaluation Scores}\label{app:raw_llm_scores}

\begin{figure}[H]
    \includegraphics[width=\textwidth]{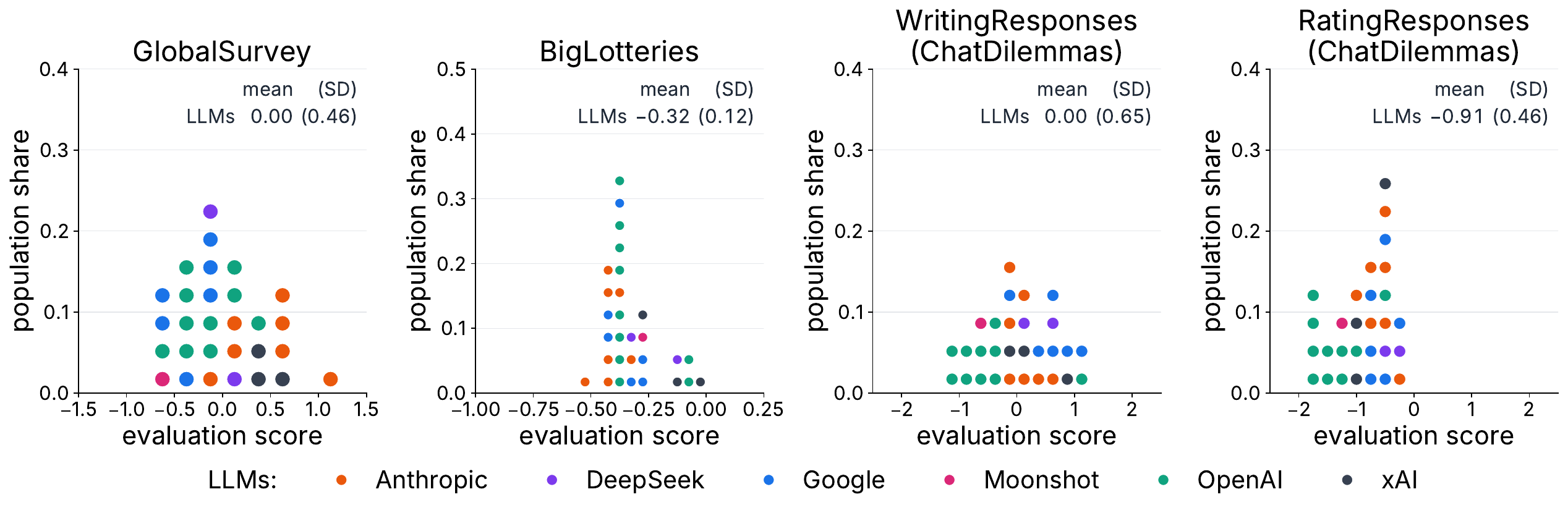}
    \caption{Distribution of the risk preference scores of the 29 LLM subjects on the four additional risk preference evaluations run on the LLMs as a robustness check (see \cref{fig:score_distributions} for the score distributions of the other evaluations).}
    \label{fig:score_distributions_llm_only}
\end{figure}

\cref{tab:llm_scores} displays the raw risk preference evaluation scores of the 29 LLMs we evaluate. See also \cref{tab:eval_scores} for summary statistics and \cref{fig:score_distributions,fig:score_distributions_llm_only} for a visualization of the score distributions.

\definecolor{scoreshade}{HTML}{3B82C4}
\begin{table}[H]
\centering
\small
\setlength{\tabcolsep}{3pt}
\begin{tabular}{@{}lwr{30pt}wr{30pt}wr{30pt}wr{30pt}wr{30pt}wr{30pt}wr{30pt}>{\columncolor{white}[\tabcolsep][0pt]}wr{30pt}@{}}
\toprule
Model & \multicolumn{2}{c}{Survey Evals} & \multicolumn{4}{c}{Dilemma Evals} & \multicolumn{2}{c}{Lottery Evals} \\
\cmidrule(lr){2-3} \cmidrule(lr){4-7} \cmidrule(l){8-9}
 & \multicolumn{1}{c}{\rotatebox{90}{\evalname{DospertSurvey}}} & \multicolumn{1}{c}{\rotatebox{90}{\evalname{GlobalSurvey}}} & \multicolumn{1}{c}{\rotatebox{90}{\shortstack[l]{\evalname{RatingResponses}\\\evalname{ChatDilemmas}}}} & \multicolumn{1}{c}{\rotatebox{90}{\shortstack[l]{\evalname{RatingResponses}\\\evalname{HumanDilemmas}}}} & \multicolumn{1}{c}{\rotatebox{90}{\shortstack[l]{\evalname{WritingResponses}\\\evalname{ChatDilemmas}}}} & \multicolumn{1}{c}{\rotatebox{90}{\shortstack[l]{\evalname{WritingResponses}\\\evalname{HumanDilemmas}}}} & \multicolumn{1}{c}{\rotatebox{90}{\evalname{SmallLotteries}}} & \multicolumn{1}{c@{}}{\rotatebox{90}{\evalname{BigLotteries}}} \\
\midrule
Grok 4.20 & \cellcolor{scoreshade!55}3.34 & \cellcolor{scoreshade!32}0.34 & \cellcolor{scoreshade!51}$-0.38$ & \cellcolor{scoreshade!39}$-1.09$ & \cellcolor{scoreshade!45}0.81 & \cellcolor{scoreshade!50}0.53 & \cellcolor{scoreshade!28}$-0.013$ & \cellcolor{scoreshade!45}$-0.13$ \\
Grok 4.5 & \cellcolor{scoreshade!48}3.19 & \cellcolor{scoreshade!37}0.51 & \cellcolor{scoreshade!27}$-1.00$ & \cellcolor{scoreshade!25}$-1.36$ & \cellcolor{scoreshade!28}0.07 & \cellcolor{scoreshade!28}$-0.03$ & \cellcolor{scoreshade!25}$-0.017$ & \cellcolor{scoreshade!31}$-0.25$ \\
Claude Opus 4.7 & \cellcolor{scoreshade!47}3.16 & \cellcolor{scoreshade!40}0.63 & \cellcolor{scoreshade!48}$-0.46$ & \cellcolor{scoreshade!46}$-0.95$ & \cellcolor{scoreshade!31}0.21 & \cellcolor{scoreshade!39}0.26 & \cellcolor{scoreshade!0}$-0.053$ & \cellcolor{scoreshade!13}$-0.40$ \\
Claude Opus 5 & \cellcolor{scoreshade!44}3.08 & \cellcolor{scoreshade!55}1.12 & \cellcolor{scoreshade!34}$-0.83$ & \cellcolor{scoreshade!45}$-0.99$ & \cellcolor{scoreshade!21}$-0.18$ & \cellcolor{scoreshade!39}0.25 & \cellcolor{scoreshade!7}$-0.043$ & \cellcolor{scoreshade!9}$-0.43$ \\
Claude Opus 4.6 & \cellcolor{scoreshade!38}2.94 & \cellcolor{scoreshade!40}0.62 & \cellcolor{scoreshade!52}$-0.37$ & \cellcolor{scoreshade!50}$-0.88$ & \cellcolor{scoreshade!39}0.55 & \cellcolor{scoreshade!47}0.45 & \cellcolor{scoreshade!25}$-0.017$ & \cellcolor{scoreshade!13}$-0.40$ \\
Grok 4.3 & \cellcolor{scoreshade!38}2.94 & \cellcolor{scoreshade!31}0.30 & \cellcolor{scoreshade!28}$-0.98$ & \cellcolor{scoreshade!20}$-1.46$ & \cellcolor{scoreshade!23}$-0.10$ & \cellcolor{scoreshade!20}$-0.23$ & \cellcolor{scoreshade!25}$-0.017$ & \cellcolor{scoreshade!55}$-0.05$ \\
Claude Opus 4.5 & \cellcolor{scoreshade!37}2.93 & \cellcolor{scoreshade!27}0.18 & \cellcolor{scoreshade!51}$-0.39$ & \cellcolor{scoreshade!55}$-0.79$ & \cellcolor{scoreshade!33}0.29 & \cellcolor{scoreshade!45}0.41 & \cellcolor{scoreshade!14}$-0.033$ & \cellcolor{scoreshade!15}$-0.38$ \\
Claude Sonnet 4.5 & \cellcolor{scoreshade!37}2.93 & \cellcolor{scoreshade!15}$-0.23$ & \cellcolor{scoreshade!39}$-0.70$ & \cellcolor{scoreshade!43}$-1.03$ & \cellcolor{scoreshade!25}$-0.03$ & \cellcolor{scoreshade!46}0.43 & \cellcolor{scoreshade!23}$-0.020$ & \cellcolor{scoreshade!0}$-0.51$ \\
Claude Sonnet 4.6 & \cellcolor{scoreshade!36}2.90 & \cellcolor{scoreshade!26}0.13 & \cellcolor{scoreshade!45}$-0.54$ & \cellcolor{scoreshade!50}$-0.88$ & \cellcolor{scoreshade!26}0.02 & \cellcolor{scoreshade!30}0.01 & \cellcolor{scoreshade!28}$-0.013$ & \cellcolor{scoreshade!24}$-0.31$ \\
GPT-5.1 & \cellcolor{scoreshade!35}2.89 & \cellcolor{scoreshade!36}0.48 & \cellcolor{scoreshade!17}$-1.26$ & \cellcolor{scoreshade!17}$-1.51$ & \cellcolor{scoreshade!18}$-0.34$ & \cellcolor{scoreshade!23}$-0.16$ & \cellcolor{scoreshade!23}$-0.020$ & \cellcolor{scoreshade!16}$-0.37$ \\
Claude Opus 4.8 & \cellcolor{scoreshade!33}2.84 & \cellcolor{scoreshade!42}0.67 & \cellcolor{scoreshade!32}$-0.88$ & \cellcolor{scoreshade!46}$-0.97$ & \cellcolor{scoreshade!24}$-0.09$ & \cellcolor{scoreshade!35}0.15 & \cellcolor{scoreshade!25}$-0.017$ & \cellcolor{scoreshade!9}$-0.44$ \\
GPT-5 & \cellcolor{scoreshade!26}2.67 & \cellcolor{scoreshade!28}0.21 & \cellcolor{scoreshade!13}$-1.35$ & \cellcolor{scoreshade!7}$-1.71$ & \cellcolor{scoreshade!19}$-0.27$ & \cellcolor{scoreshade!27}$-0.06$ & \cellcolor{scoreshade!25}$-0.017$ & \cellcolor{scoreshade!17}$-0.37$ \\
GPT-5.4 nano & \cellcolor{scoreshade!26}2.66 & \cellcolor{scoreshade!17}$-0.16$ & \cellcolor{scoreshade!28}$-0.98$ & \cellcolor{scoreshade!30}$-1.27$ & \cellcolor{scoreshade!3}$-0.93$ & \cellcolor{scoreshade!24}$-0.14$ & \cellcolor{scoreshade!28}$-0.013$ & \cellcolor{scoreshade!52}$-0.07$ \\
Gemini 3 Flash Preview & \cellcolor{scoreshade!25}2.64 & \cellcolor{scoreshade!21}$-0.04$ & \cellcolor{scoreshade!34}$-0.83$ & \cellcolor{scoreshade!20}$-1.45$ & \cellcolor{scoreshade!55}1.20 & \cellcolor{scoreshade!52}0.58 & \cellcolor{scoreshade!25}$-0.017$ & \cellcolor{scoreshade!30}$-0.26$ \\
GPT-5.4 & \cellcolor{scoreshade!23}2.61 & \cellcolor{scoreshade!12}$-0.31$ & \cellcolor{scoreshade!3}$-1.62$ & \cellcolor{scoreshade!0}$-1.84$ & \cellcolor{scoreshade!10}$-0.66$ & \cellcolor{scoreshade!11}$-0.45$ & \cellcolor{scoreshade!25}$-0.017$ & \cellcolor{scoreshade!19}$-0.35$ \\
GPT-5.6 sol & \cellcolor{scoreshade!22}2.57 & \cellcolor{scoreshade!29}0.24 & \cellcolor{scoreshade!0}$-1.69$ & \cellcolor{scoreshade!5}$-1.75$ & \cellcolor{scoreshade!0}$-1.06$ & \cellcolor{scoreshade!0}$-0.74$ & \cellcolor{scoreshade!25}$-0.017$ & \cellcolor{scoreshade!15}$-0.39$ \\
GPT-5.6 terra & \cellcolor{scoreshade!21}2.54 & \cellcolor{scoreshade!1}$-0.69$ & \cellcolor{scoreshade!1}$-1.65$ & \cellcolor{scoreshade!21}$-1.43$ & \cellcolor{scoreshade!0}$-1.05$ & \cellcolor{scoreshade!7}$-0.56$ & \cellcolor{scoreshade!25}$-0.017$ & \cellcolor{scoreshade!15}$-0.39$ \\
DeepSeek V4 Flash (Preview) & \cellcolor{scoreshade!20}2.53 & \cellcolor{scoreshade!21}$-0.03$ & \cellcolor{scoreshade!53}$-0.34$ & \cellcolor{scoreshade!35}$-1.18$ & \cellcolor{scoreshade!30}0.18 & \cellcolor{scoreshade!31}0.05 & \cellcolor{scoreshade!55}0.027 & \cellcolor{scoreshade!25}$-0.30$ \\
GPT-5.5 & \cellcolor{scoreshade!19}2.51 & \cellcolor{scoreshade!14}$-0.26$ & \cellcolor{scoreshade!1}$-1.65$ & \cellcolor{scoreshade!1}$-1.83$ & \cellcolor{scoreshade!15}$-0.44$ & \cellcolor{scoreshade!12}$-0.44$ & \cellcolor{scoreshade!25}$-0.017$ & \cellcolor{scoreshade!17}$-0.37$ \\
Gemini 3.1 Pro Preview & \cellcolor{scoreshade!19}2.51 & \cellcolor{scoreshade!7}$-0.49$ & \cellcolor{scoreshade!42}$-0.61$ & \cellcolor{scoreshade!26}$-1.34$ & \cellcolor{scoreshade!50}0.98 & \cellcolor{scoreshade!27}$-0.06$ & \cellcolor{scoreshade!14}$-0.033$ & \cellcolor{scoreshade!19}$-0.35$ \\
DeepSeek V4 Pro (Preview) & \cellcolor{scoreshade!19}2.50 & \cellcolor{scoreshade!23}0.04 & \cellcolor{scoreshade!44}$-0.56$ & \cellcolor{scoreshade!17}$-1.52$ & \cellcolor{scoreshade!40}0.60 & \cellcolor{scoreshade!31}0.06 & \cellcolor{scoreshade!34}$-0.004$ & \cellcolor{scoreshade!45}$-0.13$ \\
GPT-5.2 & \cellcolor{scoreshade!19}2.49 & \cellcolor{scoreshade!17}$-0.17$ & \cellcolor{scoreshade!2}$-1.64$ & \cellcolor{scoreshade!2}$-1.81$ & \cellcolor{scoreshade!4}$-0.92$ & \cellcolor{scoreshade!21}$-0.21$ & \cellcolor{scoreshade!30}$-0.010$ & \cellcolor{scoreshade!14}$-0.39$ \\
GPT-4.1 & \cellcolor{scoreshade!17}2.46 & \cellcolor{scoreshade!7}$-0.48$ & \cellcolor{scoreshade!46}$-0.52$ & \cellcolor{scoreshade!23}$-1.41$ & \cellcolor{scoreshade!51}1.02 & \cellcolor{scoreshade!55}0.65 & \cellcolor{scoreshade!14}$-0.033$ & \cellcolor{scoreshade!14}$-0.40$ \\
Gemini 3.5 Flash & \cellcolor{scoreshade!15}2.42 & \cellcolor{scoreshade!18}$-0.13$ & \cellcolor{scoreshade!34}$-0.83$ & \cellcolor{scoreshade!19}$-1.47$ & \cellcolor{scoreshade!33}0.31 & \cellcolor{scoreshade!32}0.07 & \cellcolor{scoreshade!25}$-0.017$ & \cellcolor{scoreshade!20}$-0.34$ \\
Gemini 3.6 Flash & \cellcolor{scoreshade!14}2.39 & \cellcolor{scoreshade!17}$-0.14$ & \cellcolor{scoreshade!33}$-0.85$ & \cellcolor{scoreshade!24}$-1.38$ & \cellcolor{scoreshade!43}0.71 & \cellcolor{scoreshade!35}0.14 & \cellcolor{scoreshade!25}$-0.017$ & \cellcolor{scoreshade!28}$-0.27$ \\
GPT-5.4 mini & \cellcolor{scoreshade!13}2.37 & \cellcolor{scoreshade!11}$-0.35$ & \cellcolor{scoreshade!11}$-1.41$ & \cellcolor{scoreshade!7}$-1.70$ & \cellcolor{scoreshade!8}$-0.75$ & \cellcolor{scoreshade!2}$-0.68$ & \cellcolor{scoreshade!7}$-0.043$ & \cellcolor{scoreshade!53}$-0.06$ \\
Gemini 2.5 Pro & \cellcolor{scoreshade!9}2.27 & \cellcolor{scoreshade!3}$-0.61$ & \cellcolor{scoreshade!48}$-0.45$ & \cellcolor{scoreshade!21}$-1.43$ & \cellcolor{scoreshade!40}0.60 & \cellcolor{scoreshade!31}0.04 & \cellcolor{scoreshade!25}$-0.017$ & \cellcolor{scoreshade!11}$-0.42$ \\
Kimi K2.7-Code & \cellcolor{scoreshade!6}2.20 & \cellcolor{scoreshade!0}$-0.73$ & \cellcolor{scoreshade!17}$-1.25$ & \cellcolor{scoreshade!11}$-1.63$ & \cellcolor{scoreshade!10}$-0.64$ & \cellcolor{scoreshade!18}$-0.27$ & \cellcolor{scoreshade!53}0.024 & \cellcolor{scoreshade!30}$-0.26$ \\
Gemini 2.5 Flash & \cellcolor{scoreshade!0}2.06 & \cellcolor{scoreshade!2}$-0.65$ & \cellcolor{scoreshade!55}$-0.28$ & \cellcolor{scoreshade!35}$-1.17$ & \cellcolor{scoreshade!24}$-0.08$ & \cellcolor{scoreshade!27}$-0.05$ & \cellcolor{scoreshade!25}$-0.017$ & \cellcolor{scoreshade!12}$-0.41$ \\
\bottomrule
\end{tabular}
\caption{Scores of the 29 LLMs on the 8 risk preference evaluations, sorted by \evalname{DospertSurvey} score.}
\label{tab:llm_scores}
\end{table}

\begin{table}[H]
\centering
\setlength{\tabcolsep}{4pt}
\begin{tabular}{@{}lrrrr@{\hskip 22pt}rrrrr@{}}
\toprule
 & \multicolumn{4}{c}{LLMs ($n = 29$)} & \multicolumn{5}{c}{Human participants} \\
\cmidrule(r{22pt}){2-5} \cmidrule{6-10}
Evaluation & Mean & SD & Min & Max & $n$ & Mean & SD & Min & Max \\
\midrule
\multicolumn{10}{@{}l}{\emph{Survey evaluations}} \\
\evalname{DospertSurvey} & 2.67 & 0.31 & 2.06 & 3.34 & 547 & 3.03 & 0.72 & 1.50 & 5.40 \\
\evalname{GlobalSurvey} & 0.00 & 0.46 & $-0.73$ & 1.12 & -- & -- & -- & -- & -- \\
\addlinespace
\multicolumn{10}{@{}l}{\emph{Lottery evaluations}} \\
\evalname{SmallLotteries} & $-0.018$ & 0.016 & $-0.053$ & 0.027 & 547 & $-0.041$ & 0.159 & $-0.450$ & 0.450 \\
\evalname{BigLotteries} & $-0.32$ & 0.12 & $-0.51$ & $-0.05$ & -- & -- & -- & -- & -- \\
\addlinespace
\multicolumn{10}{@{}l}{\emph{Dilemma evaluations, \evalname{ChatDilemmas}}} \\
\evalname{WritingResponses} & 0.00 & 0.65 & $-1.06$ & 1.20 & -- & -- & -- & -- & -- \\
\evalname{RatingResponses} & $-0.91$ & 0.46 & $-1.69$ & $-0.28$ & -- & -- & -- & -- & -- \\
\addlinespace
\multicolumn{10}{@{}l}{\emph{Dilemma evaluations, \evalname{HumanDilemmas}}} \\
\evalname{WritingResponses} & 0.00 & 0.36 & $-0.74$ & 0.65 & 293 & 0.00 & 0.68 & $-1.71$ & 2.50 \\
\evalname{RatingResponses} & $-1.35$ & 0.31 & $-1.84$ & $-0.79$ & 254 & $-0.45$ & 0.96 & $-2.00$ & 2.00 \\
\bottomrule
\end{tabular}
\caption{Summary statistics of the human and LLM subjects' scores on the eight risk preference evaluations.}
\label{tab:eval_scores}
\end{table}

\clearpage

\section{{Relationship Between LLM Capabilities and Reliability}}\label{app:capability_reliability}

\begin{figure}[H]
    \includegraphics[width=\textwidth]{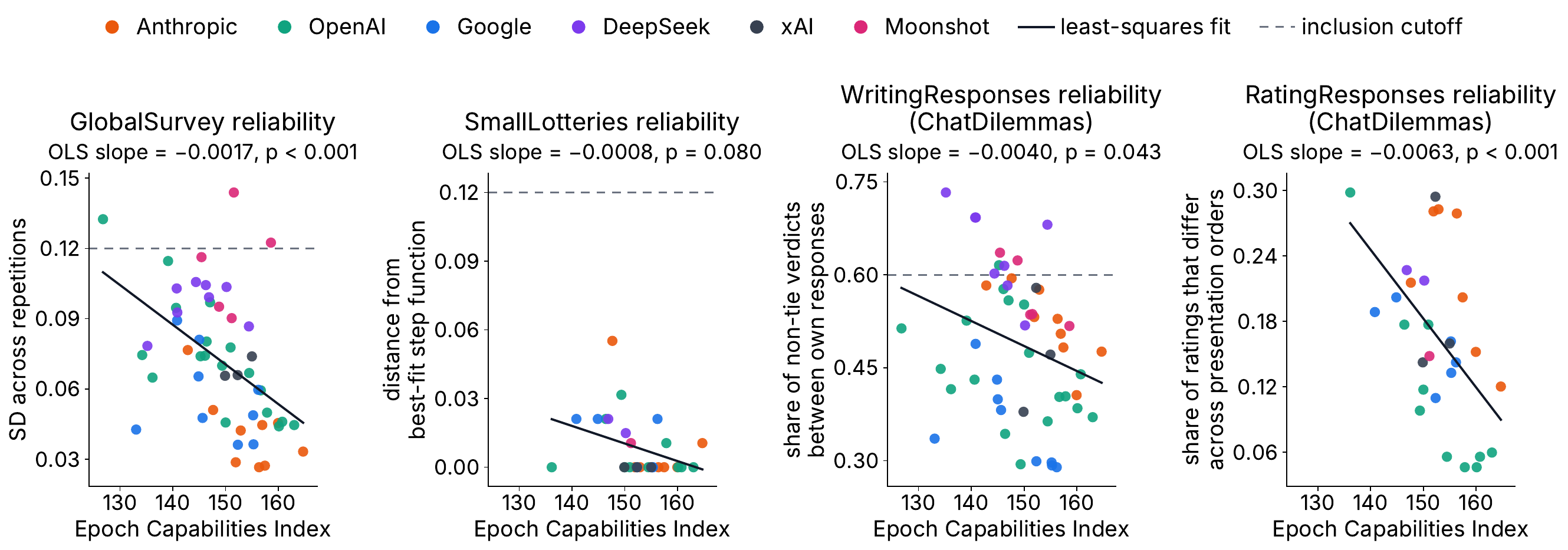}
    \caption{Relationship between LLM ECI scores and reliability scores (lower scores mean \emph{more} coherent preferences). We find that more capable LLMs have more coherent risk preferences (cf.~\cref{fig:capability_reliability}).}
    \label{fig:capability_reliability_other}
\end{figure}

{\cref{fig:capability_reliability_other} shows the relationship between LLM capabilities and reliability scores for the other four evaluations not shown in \cref{fig:capability_reliability}.}

\section{LLM Judge Validation Against Human Gold Labels}\label{app:llm_judge_validation}

The scoring of both \evalname{RatingResponses} and \evalname{WritingResponses} relies on an LLM judge (Gemini 3.8 Flash) that takes as input two responses to a dilemma, and labels which response is riskier (with ties allowed). To validate the accuracy of this LLM judge, in Experiment 1 (with \evalname{RatingResponses}), we ask participants to label which of the two displayed responses is riskier, for each of the two response pairs to the two dilemmas they write. This results in $254 \cdot 2 = 508$ ground truth labeled pairs. For each such pair, we query the LLM judge 20 times (half with order flipped), aggregate by taking the median, and measure its strict (must be exact match) and weak (tie judgements count as matches to non-tie judgements) agreement rates with the human labels. The strict agreement rate is 66\% (random baseline 39\%), and the weak agreement rate is 92\% (random baseline 88\%). That is, the LLM judge performs well enough to justify using it in scoring the dilemma evaluations.

\clearpage

\section{Correlation Matrix of LLM Evaluation Scores}\label{app:correlation_matrix}

\begin{figure}[H]
    \centering
    \includegraphics[width=0.5\linewidth]{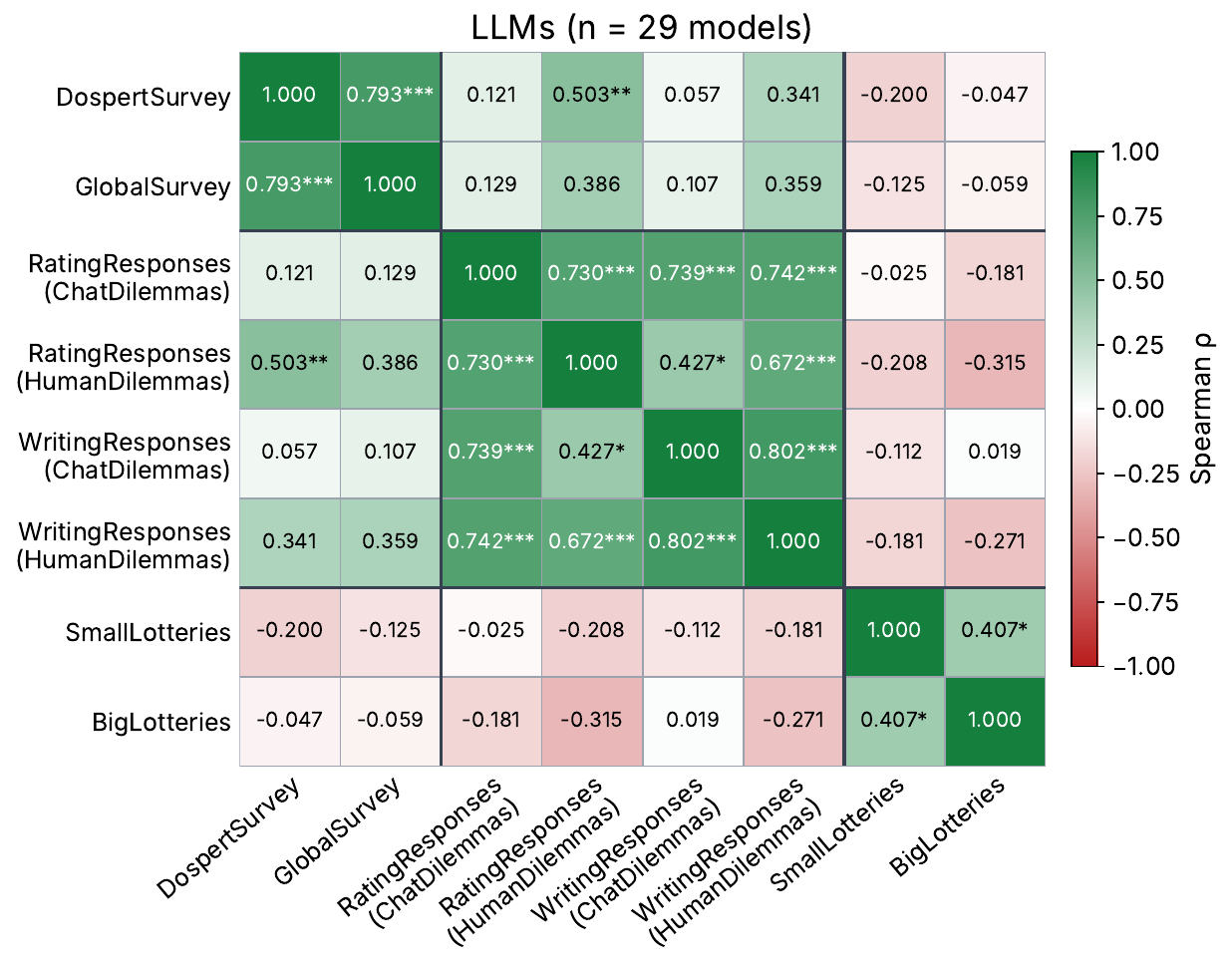}
    \caption{Correlation matrix of all eight evaluation results on LLMs. We observe (1) a block structure, where evaluations of the same type correlate the most, (2) survey evaluations exhibit positive, but weaker, correlation with dilemma evaluations, (3) lottery evaluations do not correlate with the other evaluations. *** $p < 0.01$, ** $p < 0.05$, * $p < 0.1$, Benjamini–Hochberg adjusted.}
    \label{fig:matrix}
\end{figure}

\cref{fig:matrix} displays the correlation matrix discussed in \cref{sec:correlation_llm_scores}.

\clearpage

\section{Survey vs.\ Lottery Rankings of LLMs}\label{app:survey_vs_lottery_rankings}

\begin{figure}[H]
    \centering
    \includegraphics[width=0.8\linewidth]{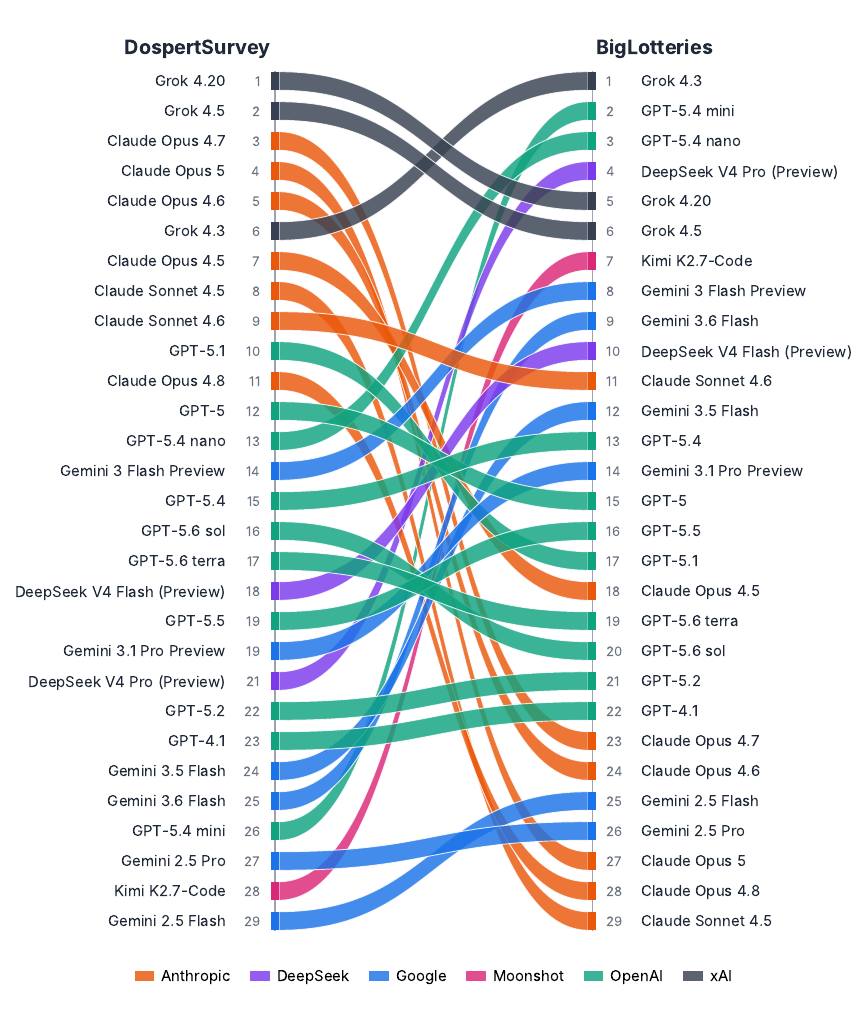}
    \caption{Ranking of the risk preference scores of the 29 LLMs we evaluate by (left) \evalname{DospertSurvey} score and (right) \evalname{BigLotteries} score. Some LLM families---like all xAI models, or pre-Gemini-3 era Google models---exhibit similar rankings under both evaluations, but other LLM families---like most Anthropic models---are relatively risk tolerant on \evalname{DospertSurvey}, but relatively risk averse on \evalname{BigLotteries}.}
    \label{fig:rank_flow}
\end{figure}

\cref{fig:rank_flow} displays the data discussed in \cref{app:diagnosing_survey_vs_lottery}.

\clearpage

\part{Additional Robustness Results}

\section{Prompt and Reasoning Level Robustness of Lottery Evaluations}\label{app:robustness_reasoning_prompt}

\begin{figure}[H]
    \includegraphics[width=\textwidth]{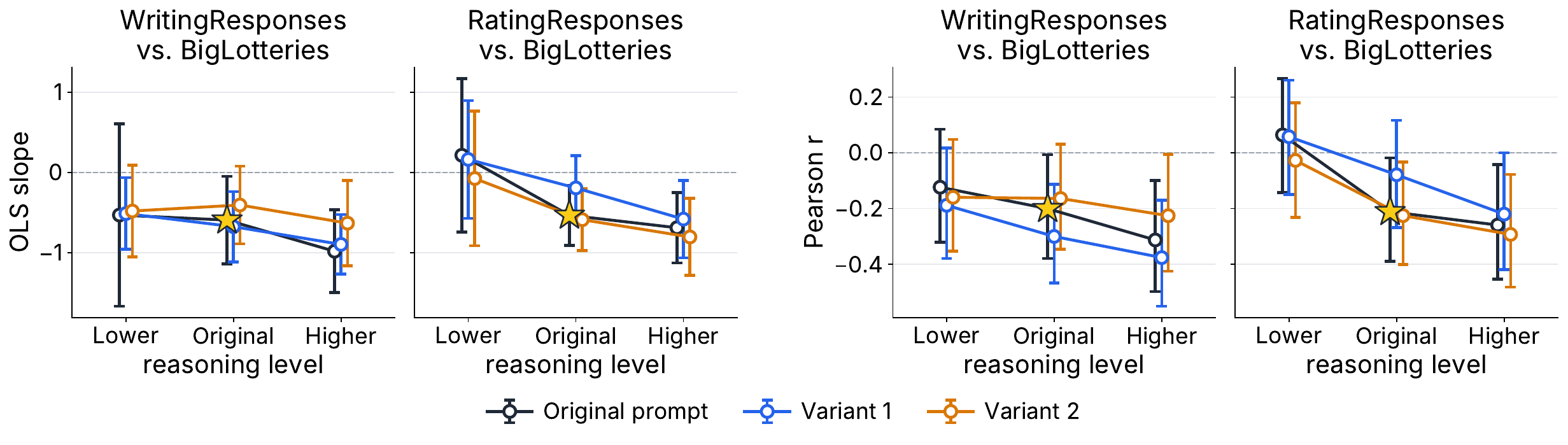}
    \caption{Relationship between dilemma evaluation scores (using \evalname{HumanDilemmas}) and \evalname{BigLotteries} scores, under three prompt variants and three reasoning levels. The left two panels display OLS slope (cf.~\cref{fig:heatmaps_slope}), the right two panels display the Pearson correlation coefficient of the fit. Error bars denote $\pm 1$ SE. We find that the main results of \cref{sec:measuring_generalizability_of_human_and_llm} are robust to changes to the prompt and reasoning level: \evalname{BigLotteries} continues to not be predictive of the dilemma evaluations.}
    \label{fig:robustness_slopes_pearson_big_lotteries_human_dilemmas}
\end{figure}

\cref{fig:robustness_slopes_pearson_big_lotteries_human_dilemmas} repeats the analysis of \cref{fig:dospert_robustness_slopes}, replacing \evalname{DospertSurvey} with \evalname{BigLotteries}. The results are qualitatively the same as in the main analysis: under (almost) no prompt variant or reasoning level is \evalname{BigLotteries} predictive of either dilemma evaluation. Interestingly, the OLS slope and Pearson correlation coefficients appear to decrease (i.e., become more negative) as inference-time compute increases. This is consistent also with \cref{sec:correlation_llm_scores}. (However, these negative slope estimates are mostly not statistically distinguishable from 0.)

\section{Statistical Robustness of the Generalizability Analysis}\label{app:robustness_generalizability}

In the main generalizability analysis (\cref{sec:main_results}), we measure generalizability of a survey or lottery evaluation to a dilemma evaluation via the slope of an OLS regression, running one separate regression for each lab-field evaluation pair. We use this analysis to conclude that survey evaluations are (mostly) predictive of dilemma evaluations, while lottery evaluations are consistently not.

In this section, we conduct two robustness checks to this statistical analysis. In \cref{app:robustness_regression}, we run regressions to predict the \evalname{RatingResponses} score from both kinds of features (survey and lottery) simultaneously. And in \cref{app:robustness_corr}, we repeat the main analysis using Spearman $\rho$ as a measure of generalizability instead. For the human data, these analyses were preregistered.

\subsection{Robustness of the OLS Regressions}\label{app:robustness_regression}

See \cref{tab:human_risk_regressions,tab:llm_risk_regressions}. For both humans and LLMs, we observe that the regression coefficients in (1) and (2)---corresponding to the values in the main analysis in \cref{sec:main_results}---are relatively stable in (3) and (4)---regressions including both features simultaneously, and optionally an interaction term. That is, lottery evaluations do not even yield predictive power in tandem with survey evaluations.

\begin{table}[H]
\centering
\caption{Predicting \evalname{RatingResponses} from \evalname{DospertSurvey} and \evalname{SmallLotteries} (human participants)}
\label{tab:human_risk_regressions}
\setlength{\tabcolsep}{5pt}
\begin{tabular}{@{}lcccc@{}}
\toprule
 & \multicolumn{4}{c@{}}{Preference for Grok 4.20 over GPT 5.6 Terra ($-2$ to $+2$)} \\
\cmidrule(l){2-5}
 & (1) & (2) & (3) & (4) \\
\midrule
\evalname{DospertSurvey} & 0.168$^{**}$ &  & 0.181$^{**}$ & 0.204$^{**}$ \\
 & (0.085) &  & (0.084) & (0.085) \\
\evalname{SmallLotteries} &  & -0.199 & -0.339 & -0.425 \\
 &  & (0.368) & (0.366) & (0.355) \\
\evalname{DospertSurvey} $\times$ \evalname{SmallLotteries} &  &  &  & -0.649 \\
 &  &  &  & (0.424) \\
Constant & -0.448$^{***}$ & -0.448$^{***}$ & -0.448$^{***}$ & -0.434$^{***}$ \\
 & (0.060) & (0.060) & (0.060) & (0.059) \\
\midrule
Observations & 254 & 254 & 254 & 254 \\
$R^2$ & 0.017 & 0.001 & 0.021 & 0.027 \\
\bottomrule
\end{tabular}
\begin{minipage}{\linewidth}
\footnotesize\emph{Notes:} OLS regression with robust standard errors predicting a participant's preference for Grok 4.20 over GPT 5.6 Terra (on a -2 to 2 Likert scale) from their \evalname{DospertSurvey} score (1 to 7 Likert scale) and/or their \evalname{SmallLotteries} score (-0.5 to 0.5 continuous scale). Both scores are normalized to mean 0.
$^{*}$ $p<0.10$, $^{**}$ $p<0.05$, $^{***}$ $p<0.01$.
\end{minipage}
\end{table}

\begin{table}[H]
\centering
\caption{Predicting \evalname{RatingResponses} from \evalname{DospertSurvey} and \evalname{SmallLotteries} (LLM subjects)}
\label{tab:llm_risk_regressions}
\setlength{\tabcolsep}{5pt}
\begin{tabular}{@{}lcccc@{}}
\toprule
 & \multicolumn{4}{c@{}}{Preference for Grok 4.20 over GPT 5.6 Terra ($-2$ to $+2$)} \\
\cmidrule(l){2-5}
 & (1) & (2) & (3) & (4) \\
\midrule
\evalname{DospertSurvey} & 0.527$^{***}$ &  & 0.510$^{***}$ & 0.488$^{**}$ \\
 & (0.165) &  & (0.181) & (0.181) \\
\evalname{SmallLotteries} &  & -4.166 & -0.980 & -1.742 \\
 &  & (3.490) & (2.691) & (2.624) \\
\evalname{DospertSurvey} $\times$ \evalname{SmallLotteries} &  &  &  & -9.386$^{*}$ \\
 &  &  &  & (5.007) \\
Constant & -1.353$^{***}$ & -1.353$^{***}$ & -1.353$^{***}$ & -1.368$^{***}$ \\
 & (0.050) & (0.057) & (0.051) & (0.054) \\
\midrule
Observations & 29 & 29 & 29 & 29 \\
$R^2$ & 0.278 & 0.047 & 0.280 & 0.307 \\
\bottomrule
\end{tabular}
\begin{minipage}{\linewidth}
\footnotesize\emph{Notes:} OLS regression with robust standard errors predicting an LLM's preference for Grok 4.20 over GPT 5.6 Terra (on a -2 to 2 Likert scale) from its \evalname{DospertSurvey} score (1 to 7 Likert scale) and/or its \evalname{SmallLotteries} score (-0.5 to 0.5 continuous scale). Both scores are normalized to mean 0.
$^{*}$ $p<0.10$, $^{**}$ $p<0.05$, $^{***}$ $p<0.01$.
\end{minipage}
\end{table}

\clearpage

\subsection{Measuring Generalizability via Spearman Rank Correlation}\label{app:robustness_corr}

\begin{figure}[H]
  \includegraphics[width=\linewidth]{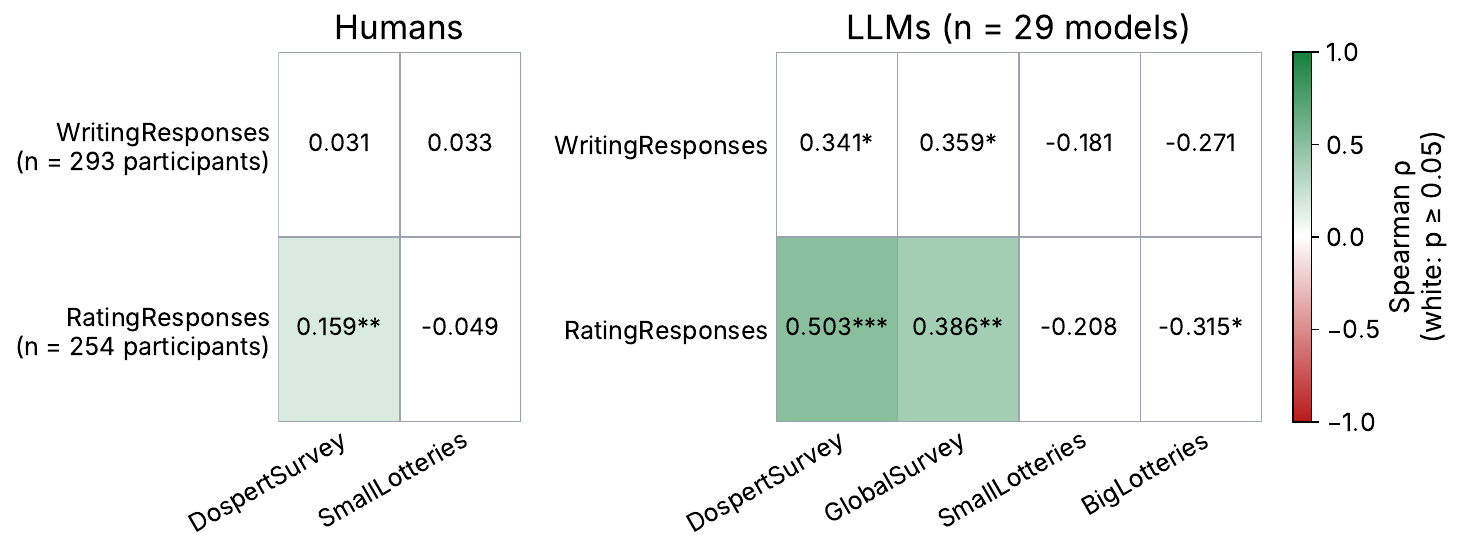}
  \caption{Relationship between dilemma evaluation scores (\evalname{WritingResponses} or \evalname{RatingResponses}, using \evalname{HumanDilemmas}) and survey or lottery evaluation scores. Each cell displays the Spearman rank correlation $\rho$ between the two evaluations. *** $p < 0.01$, ** $p < 0.05$, * $p < 0.1$.}
  \label{fig:heatmaps_spearman}
\end{figure}

\cref{fig:heatmaps_spearman} repeats the analysis of \cref{fig:heatmaps_slope}, replacing the OLS slope in each cell with the Spearman rank correlation $\rho$ between the two evaluation scores. The results are qualitatively the same as in the main analysis: survey evaluations (mostly) generalize, with a stronger correlation for LLMs than humans, while lottery evaluations do not.

\clearpage

\section{Robustness to Differences in LLMs' Writing Styles}\label{app:robustness_writing_style}

\begin{figure}[H]
    \includegraphics[width=\textwidth]{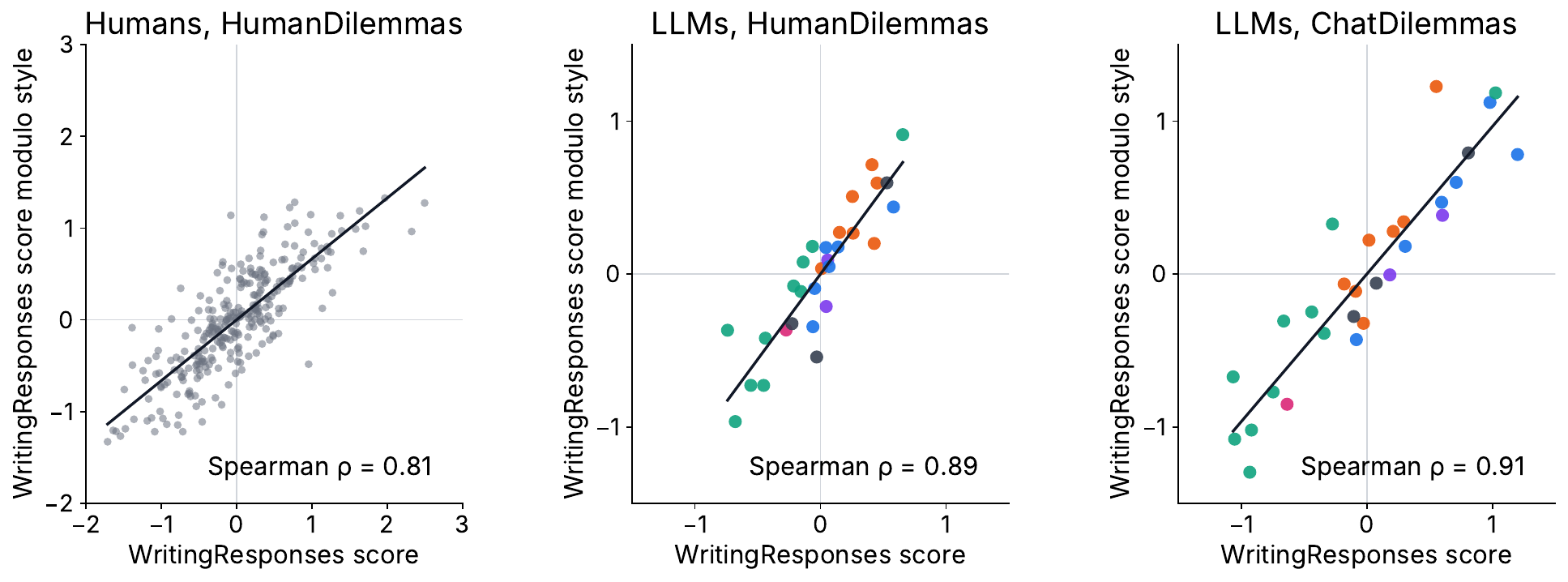}
    \caption{Relationship between \evalname{WritingResponses} scores and the same scores controlling for writing style.}
    \label{fig:scores_scatter}
\end{figure}

{\cref{fig:scores_scatter} visualizes the relationship between the standard \evalname{WritingResponses} scores and the \evalname{WritingResponses} scores modulo style (\cref{sec:writing_style_robustness}).}

\begin{figure}[H]
    \includegraphics[width=\textwidth]{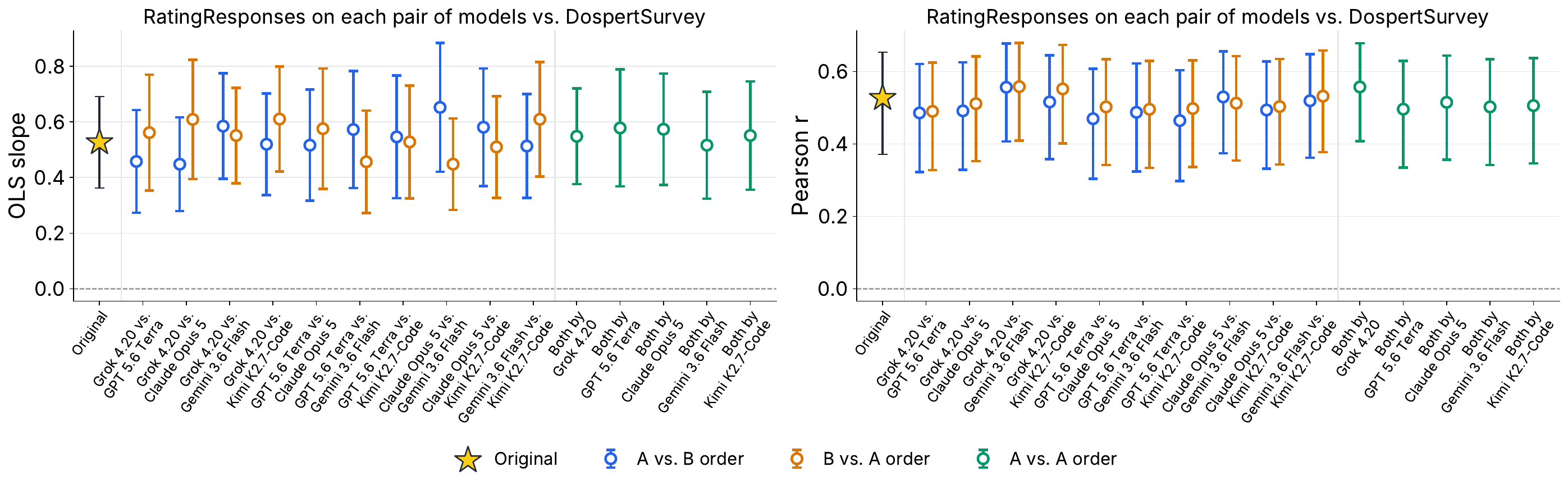}
    \caption{{Relationship between LLMs' \evalname{RatingResponses} scores and \evalname{DospertSurvey} scores, where the \evalname{RatingResponses} scores are recomputed for each pair of rewriter LLMs. The star denotes the original \evalname{RatingResponses} scores. For each pair of rewriter LLMs, ``A vs.~B order'' means the first LLM rewrote the riskier response and the second the safer response, and ``B vs.~A order'' the reverse; in ``A vs.~A order'', the same LLM rewrote both responses. The left panel displays OLS slope, the right panel displays the Pearson correlation coefficient of the fit. Error bars denote $\pm 1$ SE.}}
    \label{fig:rewrite_pairs_vs_dospert}
\end{figure}

{\cref{fig:rewrite_pairs_vs_dospert} displays the relationship between \evalname{DospertSurvey} scores and the \evalname{RatingResponses} scores recomputed for each pair of rewriter LLMs (\cref{sec:writing_style_robustness}).}

\clearpage

\part{Additional Design Details}

\section{Competency and Reliability Scoring Details}\label{app:competency_and_reliability_scoring}

\subsection{Competency scoring} We use STEER \citep{raman_steer_2024}, a suite of benchmarks designed to test the general economic decision-making capabilities of LLMs. We compute the accuracy of each LLM on the risk-related questions in STEER, from the five categories: ``completeness risk'', ``compute expected utility'', ``independence risk'', ``maximize expected utility'', ``transitivity risk''. We consider an LLM sufficiently competent if it achieves an accuracy of at least $0.95$.

\subsection{Reliability scoring} To compute reliability scores, we use evaluation-specific measures, and require that each LLM is sufficiently reliable across all evaluations.

For the survey evaluations, we normalize the scale of each question to a 0--1 scale, calculate for each LLM and each question the standard deviation of the LLM's responses (over 10 repetitions), and average over these values (for \evalname{DospertSurvey} and \evalname{GlobalSurvey} separately) to obtain two reliability scores for each LLM. We employ a cutoff of $\le 0.12$.

For the lottery evaluations, we compute two kinds of reliability scores (averaged over the five lotteries): one based on the $L^2$ distance from the best-fit step function, and one based on the $L^2$ distance from the best-fit monotone function (obtained via isotonic regression). The first score is motivated by the first question discussed in \cref{sec:introduction}, and the second score is motivated by \citet{zhu2025steeringriskpreferenceslarge}, who posit that LLMs' underlying risk preferences are probabilistic. For the first score, we employ a cutoff of $\le 0.12$, and for the second score, we employ a cutoff of $\le 0.04$.

Finally, for the dilemma evaluations, for each LLM and each dilemma, we use an LLM judge to compare pairs of the LLM's responses to that same question. For \evalname{ChatDilemmas}, we sample 20 pairs of distinct responses per LLM and question, and for \evalname{HumanDilemmas}, we judge all 3 pairs. For each pair, the LLM judge is queried twice, once with the response order flipped. For each LLM and each dataset, the reliability score is computed as the proportion of ``tie'' verdicts by the LLM judge. We employ a cutoff of $\ge 0.40$.

\clearpage

\section{Risk Preference Evaluation Dataset Summary}\label{app:risk_pref_eval_dataset_summary}

\cref{tab:eval_datasets} displays a summary of the eight risk preference evaluations we employ.

\begin{table}[H]
\centering
\setlength{\tabcolsep}{4pt}
\begin{tabular}{@{}llll@{}}
\toprule
Evaluation & Items & Score & Human participants \\
\midrule
\multicolumn{4}{@{}l}{\emph{Survey evaluations}} \\
\evalname{DospertSurvey} & 30 questions & Rating, 1 to 7 & Exp. 1 and 2 \\
\evalname{GlobalSurvey} & 53 questions & $z$-scored answer & -- \\
\addlinespace
\multicolumn{4}{@{}l}{\emph{Lottery evaluations}} \\
\evalname{SmallLotteries} & 3 lists, 33 rows & Risk premium, $(CE - EV)/\text{prize}$ & Exp. 1 and 2 \\
\evalname{BigLotteries} & 5 lists, 197 rows & Risk premium, $(CE - EV)/EV$ & -- \\
\addlinespace
\multicolumn{4}{@{}l}{\emph{Dilemma evaluations, \evalname{ChatDilemmas}}} \\
\evalname{WritingResponses} & 26 dilemmas & Bradley--Terry log-strength & -- \\
\evalname{RatingResponses} & 260 response pairs & Grok $>$ GPT preference, $-2$ to $+2$ & -- \\
\addlinespace
\multicolumn{4}{@{}l}{\emph{Dilemma evaluations, \evalname{HumanDilemmas}}} \\
\evalname{WritingResponses} & 508 dilemmas & Bradley--Terry log-strength & Exp. 2 \\
\evalname{RatingResponses} & 508 response pairs & Grok $>$ GPT preference, $-2$ to $+2$ & Exp. 1 \\
\bottomrule
\end{tabular}
\caption{Basic information about the eight risk preference evaluations we employ.}
\label{tab:eval_datasets}
\end{table}

\section{\evalname{ChatDilemmas} Filtering Details}\label{app:chat_filtering}

We employ the following semi-automatic pipeline to obtain \evalname{ChatDilemmas} from WildChat-4.8M and LMSYS-Chat-1M:
\begin{enumerate}
  \item We filter for English-language, non-toxic queries with initial user message between 40 and 3000 characters. This results in 898{,}550 remaining prompts.
  \item We embed each prompt using text-embedding-3-small. We compute the dot product of each vector with the embeddings of the following eight questions: ``Can you give me some advice?'',
``What should I do?'',
``I need advice about a decision I'm facing.'',
``Should I do it or not?'',
``What would you do if you were me?'',
``I can't decide what to do with my money.'',
``Is it a good idea or a bad idea?'', and
``I'm torn between two options, help me choose.''
We filter for prompts with cosine similarity $\ge 0.260$ with at least one of the reference questions (top 15\%). This results in 135{,}333 remaining prompts.
\item For each prompt, we use a GPT-5.6 Luna judge to assess whether it is an advice-seeking query related to risk. This results in 6{,}610 remaining prompts.
\item For each prompt, we use a Claude Sonnet 5 judge to assess whether the query (1) asks for an opinion, rather than simply information, and (2) has multiple equally reasonable potential responses. This results in 143 remaining prompts.
\item Finally, we manually reviewed this set for quality, and selected 26 prompts to include in \evalname{ChatDilemmas}.
\end{enumerate}

\clearpage

\section{LLM Inference Details}\label{app:llm_inference_details}

See \cref{tab:llm_settings} for details.

For the OpenAI, Anthropic, and Google LLMs, we query them via the provider-specific APIs. For the DeepSeek, xAI, and Moonshot LLMs, we query them via OpenRouter.

For OpenAI, Anthropic, Google, and xAI models that support reasoning, we use ``medium'' effort. For DeepSeek and Moonshot LLMs that support reasoning, we use ``high'' effort, because there is no ``medium'' option (the scale is low/high/max).

We use default temperature values for all LLMs by leaving the temperature unspecified in our API requests. (We do so because many recently-released LLMs do not support this parameter in API requests.) The temperature values listed in \cref{tab:llm_settings} reflect the defaults documented by each API provider and/or the temperature value reported in the API query's completion object.

\begin{table}[H]
\centering
\scriptsize
\renewcommand{\arraystretch}{0.9}
\setlength{\tabcolsep}{5pt}
\begin{tabular}{@{}l l c c c c c@{}}
\toprule
 & & & & & \multicolumn{2}{c@{}}{Robustness reasoning} \\
\cmidrule(l){6-7}
Model & Endpoint & Reasoning & Effort & Temperature & Lower & Higher \\
\midrule
\multicolumn{7}{@{}l}{\emph{Anthropic}} \\
\texttt{claude-sonnet-4-5-20250929} & anthropic & no & -- & 1.0 & -- & -- \\
\texttt{claude-haiku-4-5-20251001} & anthropic & no & -- & 1.0 &  &  \\
\texttt{claude-opus-4-5-20251101} & anthropic & no & medium & 1.0 & low & high \\
\texttt{claude-opus-4-6} & anthropic & yes & medium & 1.0 & low & high \\
\texttt{claude-sonnet-4-6} & anthropic & yes & medium & 1.0 & low & high \\
\texttt{claude-opus-4-7} & anthropic & yes & medium & 1.0 & low & high \\
\texttt{claude-opus-4-8} & anthropic & yes & medium & 1.0 & low & high \\
\texttt{claude-sonnet-5} & anthropic & yes & medium & 1.0 &  &  \\
\texttt{claude-opus-5} & anthropic & yes & medium & 1.0 & low & high \\
\midrule
\multicolumn{7}{@{}l}{\emph{OpenAI}} \\
\texttt{o3-mini-2025-01-31} & openai & yes & medium & 1.0 &  &  \\
\texttt{gpt-4.1-2025-04-14} & openai & no & -- & 1.0 & -- & -- \\
\texttt{gpt-4.1-mini-2025-04-14} & openai & no & -- & 1.0 &  &  \\
\texttt{gpt-4.1-nano-2025-04-14} & openai & no & -- & 1.0 &  &  \\
\texttt{o3-2025-04-16} & openai & yes & medium & 1.0 &  &  \\
\texttt{o4-mini-2025-04-16} & openai & yes & medium & 1.0 &  &  \\
\texttt{gpt-5-2025-08-07} & openai & yes & medium & 1.0 & minimal & high \\
\texttt{gpt-5-mini-2025-08-07} & openai & yes & medium & 1.0 &  &  \\
\texttt{gpt-5-nano-2025-08-07} & openai & yes & medium & 1.0 &  &  \\
\texttt{gpt-5.1-2025-11-13} & openai & yes & medium & 1.0 & none & high \\
\texttt{gpt-5.2-2025-12-11} & openai & yes & medium & 1.0 & none & high \\
\texttt{gpt-5.4-2026-03-05} & openai & yes & medium & 1.0 & none & high \\
\texttt{gpt-5.4-mini-2026-03-17} & openai & yes & medium & 1.0 & none & high \\
\texttt{gpt-5.4-nano-2026-03-17} & openai & yes & medium & 1.0 & none & high \\
\texttt{gpt-5.5-2026-04-23} & openai & yes & medium & 1.0 & none & high \\
\texttt{gpt-5.6-luna} & openai & yes & medium & 1.0 &  &  \\
\texttt{gpt-5.6-sol} & openai & yes & medium & 1.0 & none & high \\
\texttt{gpt-5.6-terra} & openai & yes & medium & 1.0 & none & high \\
\midrule
\multicolumn{7}{@{}l}{\emph{Google}} \\
\texttt{gemini-2.5-pro} & google & yes & -- & 1.0 & budget 128 & -- \\
\texttt{gemini-2.5-flash} & google & yes & -- & 1.0 & budget 0 & -- \\
\texttt{gemini-2.5-flash-lite} & google & no & -- & 1.0 &  &  \\
\texttt{gemini-3-flash-preview} & google & yes & medium & 1.0 & minimal & high \\
\texttt{gemini-3.1-pro-preview} & google & yes & medium & 1.0 & low & high \\
\texttt{gemini-3.1-flash-lite} & google & yes & medium & 1.0 &  &  \\
\texttt{gemini-3.5-flash} & google & yes & medium & 1.0 & minimal & high \\
\texttt{gemini-3.5-flash-lite} & google & yes & medium & 1.0 &  &  \\
\texttt{gemini-3.6-flash} & google & yes & medium & 1.0 & minimal & high \\
\midrule
\multicolumn{7}{@{}l}{\emph{DeepSeek}} \\
\texttt{deepseek/deepseek-chat-v3-0324} & openrouter/siliconflow & no & -- & 0.3 &  &  \\
\texttt{deepseek/deepseek-r1-0528} & openrouter/siliconflow & yes & -- & 0.6 &  &  \\
\texttt{deepseek/deepseek-chat-v3.1} & openrouter/novita & yes & -- & 0.6 &  &  \\
\texttt{deepseek/deepseek-v3.1-terminus} & openrouter/novita & yes & -- & 0.6 &  &  \\
\texttt{deepseek/deepseek-v3.2-exp} & openrouter/novita & yes & -- & 0.6 &  &  \\
\texttt{deepseek/deepseek-v3.2} & openrouter/baidu & yes & -- & -- &  &  \\
\texttt{deepseek/deepseek-v4-pro} & openrouter/baidu & yes & high & -- & none & xhigh \\
\texttt{deepseek/deepseek-v4-flash} & openrouter/deepinfra & yes & medium & 1.0 & none & xhigh \\
\texttt{deepseek/deepseek-v4-flash-0731} & openrouter/baidu & yes & medium & -- &  &  \\
\midrule
\multicolumn{7}{@{}l}{\emph{xAI}} \\
\texttt{x-ai/grok-4.20} & openrouter/xai & yes & -- & 0.7 & off & -- \\
\texttt{x-ai/grok-4.3} & openrouter/xai & yes & medium & 0.7 & none & high \\
\texttt{x-ai/grok-4.5} & openrouter/xai & yes & medium & 0.7 & low & high \\
\midrule
\multicolumn{7}{@{}l}{\emph{Moonshot}} \\
\texttt{moonshotai/kimi-k2-0905} & openrouter/novita & no & -- & -- &  &  \\
\texttt{moonshotai/kimi-k2-thinking} & openrouter/novita & yes & -- & -- &  &  \\
\texttt{moonshotai/kimi-k2.5} & openrouter/streamlake & yes & -- & -- &  &  \\
\texttt{moonshotai/kimi-k2.6} & openrouter/streamlake & yes & -- & -- &  &  \\
\texttt{moonshotai/kimi-k2.7-code} & openrouter/inceptron & yes & -- & 1.0 & -- & -- \\
\texttt{moonshotai/kimi-k3} & openrouter/moonshotai & yes & high & 1.0 &  &  \\
\bottomrule
\end{tabular}
\caption{Details on reasoning, temperature and API endpoint for the 54 LLMs evaluated.}
\label{tab:llm_settings}
\end{table}

\section{Human Experiment Design}\label{app:human_experiment_design}

In this section, we provide further details on the human experiments (\cref{sec:human_experiment_design}).

{Our preregistrations can be found at these links:
\begin{itemize}
    \item Experiment 1: https://aspredicted.org/pn6tg5.pdf
    \item  Experiment 2: https://aspredicted.org/g5jp5n.pdf
\end{itemize}}

{For a page-by-page walkthrough of Experiments 1 and 2, see \url{https://github.com/sara-fish/llm-risk-preferences/blob/main/human_experiment_walkthrough.pdf}.}

\subsection{Participants}

For Experiment 1, 420 participants were recruited via Prolific on September 21, 2026. For Experiment 2, 420 (distinct) participants were recruited via Prolific on September 24, 2026. Participants were compensated at an average rate of \$18/hr. In accordance with Prolific's guidelines\footnote{\url{https://researcher-help.prolific.com/en/articles/445153-prolific-s-attention-and-comprehension-check-policy}}, participants who failed two instructed item response attention checks, or the comprehension quiz, were asked to return their submission.

All participants were shown a consent form at the start of the study, and were given the opportunity to revoke consent at any time.

\subsubsection{Exclusion criteria}

We employ the following exclusion criteria, also detailed in our preregistration:
\begin{enumerate}
    \item First, we exclude any participant for which at least one of their written dilemmas satisfies (1) and (2), and/or (3), where the conditions are: (1) number of input events in the input box $<$ 80\% of the submitted text's characters, (2) number of pasted characters $\ge$80\% of the submitted text's characters, (3) Pangram API classifies $\ge$ 95\% of the text as AI-generated.
    \item Second, we exclude participants who fail at least one attention check. For DOSPERT we use instructed response items. For the multiple price lists, we exclude participants who make weakly dominated choices, i.e., selecting Option B in the first row or Option A in the last row. For the dilemmas, we exclude participants who answer the questions "Which response gives less risky advice?" and "Which response gives more risky advice?" inconsistently (either both answers must be "About the same", or the answers must have opposite sign).
    \item Third, we exclude participants who fail either of the two comprehension quizzes after their second attempt. A comprehension quiz is considered failed if any single question is answered incorrectly.
    \item Fourth, for Experiment 1, we exclude participants for whom any LLM responses fail to be displayed (e.g. due to an API error). For experiment 2, we exclude participants who write obviously off-topic or low-effort responses (as determined by an LLM judge), using Prolific's submission rejection criteria.
\end{enumerate}

After applying these exclusion criteria, we were left with 254 participants in Experiment 1 (60\% inclusion rate) and 293 participants in Experiment 2 (70\% inclusion rate).

\subsubsection{Demographics}

\begin{figure}[H]
    \centering
    \includegraphics[width=\linewidth]{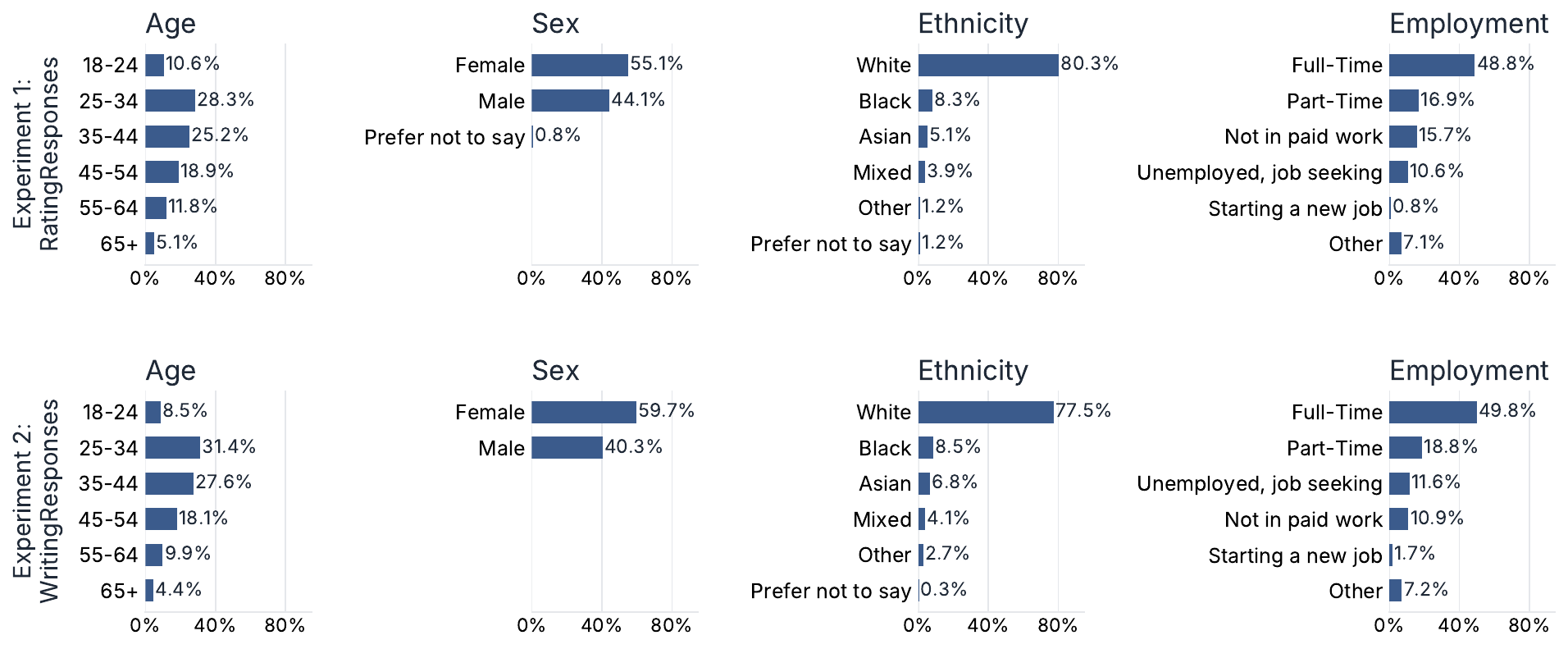}
    \caption{Age, sex, ethnicity, and employment information of the participants in Experiment 1 (254 participants) and Experiment 2 (293 participants).}
    \label{fig:demographics}
\end{figure}

We recruited participants using Prolific's standard sampling methods, filtering for participants with first language English and current country of residence in the United States or United Kingdom. See \cref{fig:demographics} for statistics on age, sex, ethnicity, and employment.

\subsection{Design}

Experiment 1 consists of three modules: \evalname{DospertSurvey}, \evalname{SmallLotteries}, and \evalname{RatingResponses}. Experiment 2 consists of three modules: \evalname{DospertSurvey}, \evalname{SmallLotteries}, and \evalname{WritingResponses}. For each experiment, the responses module is randomized to appear either first or last, and the other two modules appear in a random order. Below we describe each of the four modules in greater detail. {See \cref{fig:study_design} for a schematic.}

\subsubsection{\evalname{DospertSurvey}}

Participants are asked to complete the 30 DOSPERT survey questions (answering on a 7-point Likert scale), spread over 3 pages. Two instructed item response questions are included as attention checks.

\subsubsection{\evalname{SmallLotteries}}

This module begins with several pages of training materials, explaining to participants the following facts about the experiment:
\begin{itemize}
    \item They will be evaluating lotteries, such as: {``You might {get \$20} with a {50\% chance}, otherwise you get nothing.''}
    \item They will be presented with choices between a risky lottery (left-hand side of the multiple price list, see \cref{fig:study_design} middle panel) and a safe payment (right-hand side of the multiple price list), and asked to make a selection in each row.
    \item They have a 1 in 10 chance of being paid a bonus based on their choice in a randomly selected row.
\end{itemize}

Participants are then given a brief comprehension check, and screened out of the experiment after two attempts. Finally, participants who pass are asked to complete the three multiple price lists from \evalname{SmallLotteries}, displayed in a random order.

\subsubsection{\evalname{RatingResponses}}

In the first part of the module, participants are asked to \emph{write} two risky dilemmas from their own life. First, participants read through training materials explaining desiderata for the risky dilemmas. Then, they are given a brief comprehension check (simpler than the \evalname{SmallLotteries} check). Finally, they are asked to write their two dilemmas.

In the second part of the module, participants are asked to read and \emph{rate} two LLM-generated responses to their dilemma (generated by Grok 4.20 and GPT 5.6 Terra, presented unlabeled in random order). Participants are asked to answer the following questions (all on 5-point scale, presented in random order):
\begin{itemize}
    \item Which response do you prefer?
    \item Which response gives more risky advice?
    \item Which response would you want to hear?
    \item Which response gives less risky advice?
\end{itemize}

The two more/less risky questions are used as an attention check, and as gold labels to validate the LLM judge (see \cref{app:llm_judge_validation}). The other two questions are averaged to quantify the participant's degree of preference of Grok 4.20 over GPT 5.6 Terra.

\subsubsection{\evalname{WritingResponses}}

Participants are asked to \emph{respond to} four risky dilemmas written by other people (the participants in Experiment 1). First, participants read through training materials explaining desiderata for responses. Then, they are given a brief comprehension check (simpler than the \evalname{SmallLotteries} check). Finally, they are asked to write their four responses.

\section{Prompts}\label{app:prompts}

For a markdown file listing all of the prompts, see \url{https://github.com/sara-fish/llm-risk-preferences/blob/main/prompts.md}.

\end{document}